\documentclass[aps,twocolumn,superscriptaddress,aps, floatfix]{revtex4}
\usepackage{graphicx}
\usepackage{color}
\newcommand{\be}{\begin{equation}}
\newcommand{\ee}{\end{equation}}
\newcommand{\bea}{\begin{eqnarray}}
\newcommand{\eea}{\end{eqnarray}}
\newcommand{\kca}{KCo$_2$As$_2$}

\begin{document}
\title{KCo$_2$As$_2$: A New Portal for the Physics of High-Purity Metals}
\author{Abhishek Pandey}
\email{abhishek.pandey@wits.ac.za}
\affiliation {Ames Laboratory, Iowa State University, Ames, Iowa 50011, USA}
\affiliation{Materials Physics Research Institute, School of Physics, University of the Witwatersrand, Johannesburg, Gauteng 2050, South Africa}
\author{Y. Liu}
\affiliation {Ames Laboratory, Iowa State University, Ames, Iowa 50011, USA}
\affiliation{Present Address: Crystal Growth Facility, Institute of Physics, \'Ecole Polytechnique F\'ed\'erale de Lausanne, Lausanne CH-1015, Switzerland}
\author{Saroj L. Samal} 
\affiliation {Ames Laboratory, Iowa State University, Ames, Iowa 50011, USA}
\affiliation{Department of Chemistry, National Institute of Technology Rourkela, Rourkela 769008, India} 
\author{Yevhen Kushnirenko} 
\affiliation {Ames Laboratory, Iowa State University, Ames, Iowa 50011, USA}
\author{A. Kaminski}
\affiliation {Ames Laboratory, Iowa State University, Ames, Iowa 50011, USA}
\affiliation {Department of Physics and Astronomy, Iowa State University, Ames, Iowa 50011, USA}
\author{D. J. Singh}
\affiliation{Department of Physics and Astronomy, University of Missouri, Columbia, Missouri 65211, USA}
\author{D. C. Johnston}
\email{johnston@ameslab.gov}
\affiliation {Ames Laboratory, Iowa State University, Ames, Iowa 50011, USA}
\affiliation {Department of Physics and Astronomy, Iowa State University, Ames, Iowa 50011, USA}

\date{\today}

\begin{abstract}

High-quality single crystals of KCo$_2$As$_2$ with the body-centered tetragonal ThCr$_2$Si$_2$ structure were grown using KAs self flux. Structural, magnetic, thermal, and electrical transport were investigated.  No clear evidence for any phase transitions was found in the temperature range 2--300~K\@.  The in-plane electrical resistivity $\rho$ versus temperature $T$ is highly unusual, showing a $T^4$ behavior below 30~K and an anomalous positive curvature up to 300~K which is different from the linear behavior expected from the Bloch-Gr\"uneisen theory for electron scattering by acoustic phonons.  This positive curvature has been previously observed in the in-plane resistivity of high-conductivity layered delafossites such as PdCoO$_2$ and PtCoO$_2$.  The in-plane $\rho(T\to0) = 0.36~\mu\Omega$\,cm of KCo$_2$As$_2$ is exceptionally small for this class of compounds.  The material also exhibits a nearly linear magnetoresistance at low~$T$ which attains a value of about 40\% at $T=2$~K and magnetic field $H= 80$~kOe.  The magnetic susceptibility $\chi$ of KCo$_2$As$_2$ is isotropic and about an order of magnitude smaller than the values for the related compounds SrCo$_2$As$_2$ and BaCo$_2$As$_2$.  The $\chi$ increases above 100~K which is found from our first-principles calculations to arise from a sharp peak in the electronic density of states just above the Fermi energy~$E_{\rm F}$.  Heat capacity $C_{\rm p}(T)$ data at low~$T$ yield an electronic  density of states $N(E_{\rm F})$ that is about 36\% larger than predicted by the first-principles theory.  The $C_{\rm p}(T)$ data near room temperature suggest the presence of excited optic vibration modes which may also be the source of the positive curvature in $\rho(T)$.  Angle-resolved photoemission spectroscopy measurements are compared with the theoretical predictions of the band structure and Fermi surfaces.  Our results show that KCo$_2$As$_2$ provides a new avenue for investigating the physics of high-purity metals.  

\end{abstract}

\maketitle

\section{Introduction}

High-temperature superconductivity in iron-based layered pnictides and chalcogenides was discovered 14 years ago. These materials crystallize in tetragonal structures at room temperature~\cite{Kamihara-2008, Wang-2008, Paglione-2010, Johnston-2010, Stewart-2011, Fernandes2022}. It is thus of interest to investigate isostructural materials containing other $3d$ transition metals. For example, CaCo$_2$As$_2$ was found to exhibit long-range antiferromagnetic (AFM) order~\cite{Anand-2014d}, BaCo$_2$As$_2$ exhibits no magnetic ordering \cite{Anand-2014a}, whereas SrCo$_2$As$_2$ manifests a ground state that carries precursors to magnetic ordering in the form of AFM  fluctuations but also exhibits ferromagnetic (FM) fluctuations~\cite{Pandey-2013b, Jayasekara-2013,Li-2019}.  In these cases the formal oxidation state of the Co ions is 2+.  Here we study KCo$_2$As$_2$ with the layered body-centered tetragonal (bct) ThCr$_2$Si$_2$ structure. This is the structure of the 122-type iron-arsenide superconductors. Additionally, the oxidation state of 1+ associated with the K ion is lower than the value of 2+ associated with the alkaline-earth Ca, Sr, and Ba ions. This results in an extra hole in the (CoAs)$_2$ sublattice, which corresponds to an oxidation state of 2.5+ for the Co ions. This half-integer fractional Co oxidation state makes this compound quite promising in a search for novel properties. 

Our investigation reveals a nonmagnetic ground state for \kca.  The magnetic susceptibility versus temperature $\chi(T)$ data are isotropic with a value which is one order of magnitude smaller than the $\chi$ of the other analogs SrCo$_2$As$_2$ and BaCo$_2$As$_2$ that also exhibit no evidence for any type of long-range ordering. On the other hand, we discovered interesting electronic-transport properties of this compound, such as the observation of an unexpected $T^4$ behavior at low temperatures and anomalous positive curvature of resistivity at high temperatures. These results are unparalleled in this class of materials. Analysis of transport and thermodynamic properties in relation to first principles calculations suggests nontrivial correlated electron behavior while at the same time maintaining high electrical conductivity.

The experimental and theoretical details of our studies are presented in Sec.~\ref{Details}. The results of our single-crystal x-ray diffraction (XRD) measurements are given in Sec.~\ref{XRD}.  The measurements of electronic transport, heat capacity,  and magnetic properties are presented in Secs.~\ref{ElectronicTransport}, \ref{Cp}, and \ref{Magnetism}, respectively.   The theoretical studies of the density of states, Fermi surfaces, and electronic transport are given in Sec.~\ref{Theory} and the Angle-resolved photoemission spectroscopy (ARPES) study of the band structure in Sec.~\ref{ARPES}.  A discussion of the results is given in Sec.~\ref{Discussion} and concluding remarks are presented in Sec.~\ref{Conclusion}.

\section{\label{Details} Experimental and Theoretical Details} 

Single crystals of KCo$_2$As$_2$ were grown using solution growth technique using KAs self flux. High-purity elements K ($99.99\%$), Co ($99.99\%$) and As ($99.99\%$) were taken in a molar ratio of 5:2:6 in an alumina crucible, which was sealed inside a Ta tube. This assembly was then sealed inside an evacuated quartz tube, put into a vertical tube furnace and then slowly heated to 920~$^\circ$C in about 20~h, kept there for 1~h and then slowly cooled to 620~$^\circ$C in 150~h. The excess flux was decanted at this temperature. Several shiny plate-like crystals of typical size $7\times5\times0.3$~mm$^3$ were obtained. 

Several single crystals of KCo$_2$As$_2$ were sealed in capillaries inside a N$_2$-filled glovebox and single-crystal \mbox{x-ray} diffraction (XRD) data were collected at room temperature over a $2\theta$ range of $\sim6^{\circ}$ to $\sim60^{\circ}$ with $0.5^{\circ}$ scans in $\omega$ and 10~s per frame exposures using a Bruker SMART CCD diffractometer equipped with Mo-$K_\alpha$ radiation ($\lambda = 0.71073$~\AA). The APEX II program in the SMART package was used to integrate the collected reflection intensities \cite{SMART-1996}. The empirical absorption corrections were done using the SADABS program \cite{Blessing-1995}. The space group was determined with the help of XPREP and SHELXTL 6.1.3~\cite{SHELXTL-2000}. The structure was solved by direct methods and subsequently refined on $|F^{2}|$ with a combination of least-square refinements and difference Fourier maps.

The temperature $T$ and magnetic field $H$ dependence of the basal-plane electrical resistivity $\rho_{ab}(T)$ was measured using a Physical Properties Measurement System (PPMS) from Quantum Design, Inc.~(QDI). Heat capacity $C_{\rm p}$ versus $T$ measurements were also performed using the PPMS. Magnetic susceptibility $\chi$ versus $T$ and isothermal magnetization $M$ versus $H$ measurements were performed using a Magnetic Properties Measurement System from QDI\@. For these measurements, a single crystal of mass 4.319~mg was glued to a 1~mm diameter quartz rod using GE 7031 varnish. Separate measurements were performed on the quartz rod and the signal from the rod was subtracted to obtain the magnetic moment of the crystal.

ARPES data were acquired using a laboratory-based system consisting of a Scienta SES2002 electron analyzer and a GammaData helium ultraviolet lamp. All data were acquired using the HeI line with a photon energy of 21.2~eV. The angular resolution was $\sim0.13$ degree and $\sim0.5$ degree along and perpendicular to the direction of the analyzer slits, respectively. The energy resolution was set at $\sim10$~meV, confirmed by measuring the energy width between the 90\% and 10\% of the Fermi edge from the same Au reference. Custom-designed refocusing optics enabled us to accumulate high-statistics spectra in a short time without effects of sample-surface aging. The results were reproduced on several samples and on temperature cycling. 

The first-principles calculations were carried out using the Perdew-Burke-Ernzerhof generalized gradient
approximation (PBE GGA) functional \cite{pbe}. We used the general potential linearized augmented planewave (LAPW) method \cite{lapw} as implemented in WIEN2k \cite{wien2k}. The experimental lattice parameters $a = 3.813$~\AA\ and $c = 13.58$~\AA\ were used, but the internal As position was relaxed, yielding $z_{\rm As} = 0.3469$. The electronic structure and transport calculations were carried out with inclusion of spin-orbit coupling. LAPW sphere radii of 2.5 bohr were used for K and 2.1 bohr was used for Co and As. The basis set cutoff criterion was $R_{\rm min}k_{\rm max} = 9.0$, where $R_{\rm min}$ is the minimum LAPW sphere radius and $k_{\rm max}$ is the plane-wave sector cutoff.

\section{Results} 

\subsection{\label{XRD} Crystal structure}

Single-crystal XRD measurements on KCo$_2$As$_2$ reveal a bct lattice and the intensity statistics suggest a centrosymmetric space group, which was later determined to be $I4/mmm$ (No.\,139) with the help of XPREP and SHELXTL 6.1 packages \cite{SHELXTL-2000}. The refinements of the KCo$_2$As$_2$ diffraction data converged to $R_1 = 0.0574$, $R_{\rm W} = 0.1220$ for all data with a goodness of fit of 1.2 (Table~\ref{Table:Structure} in the Appendix). Our investigation confirms that this compound crystallizes in the ThCr$_2$Si$_2$-type tetragonal structure, as reported in Ref.~\cite{Rozsa-1981}. The measured crystallographic parameters $a$, $c$, $c/a$, $z_{\rm As}$ and $d_{\rm As-As}$ are listed in Table~\ref{Table:Parameters} and other refinement parameters are listed in Table~\ref{Table:Structure} in the Appendix.

\subsection{\label{ElectronicTransport} Electrical Transport}

\begin{figure}
\includegraphics[width=3in]{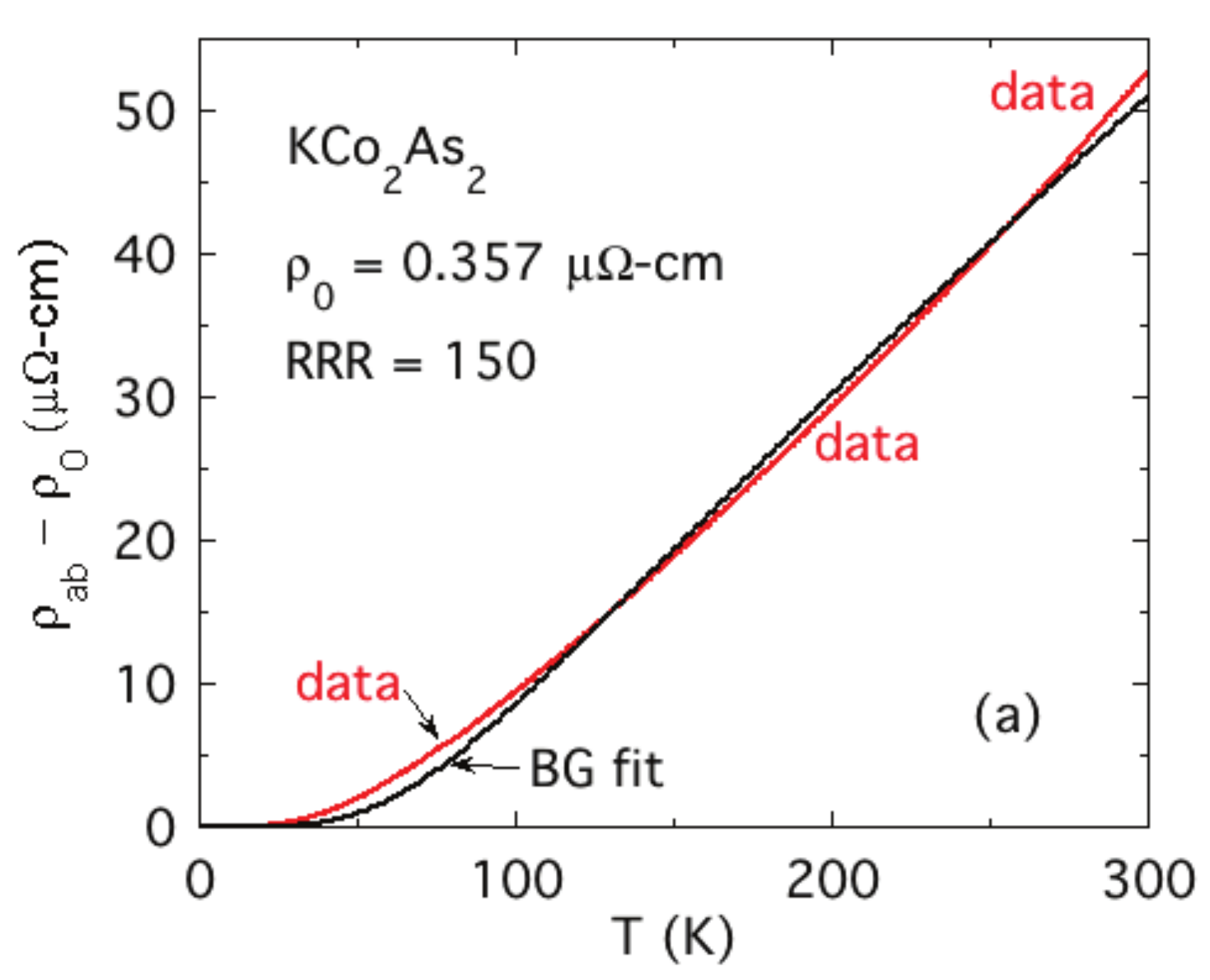}
\includegraphics[width=3in]{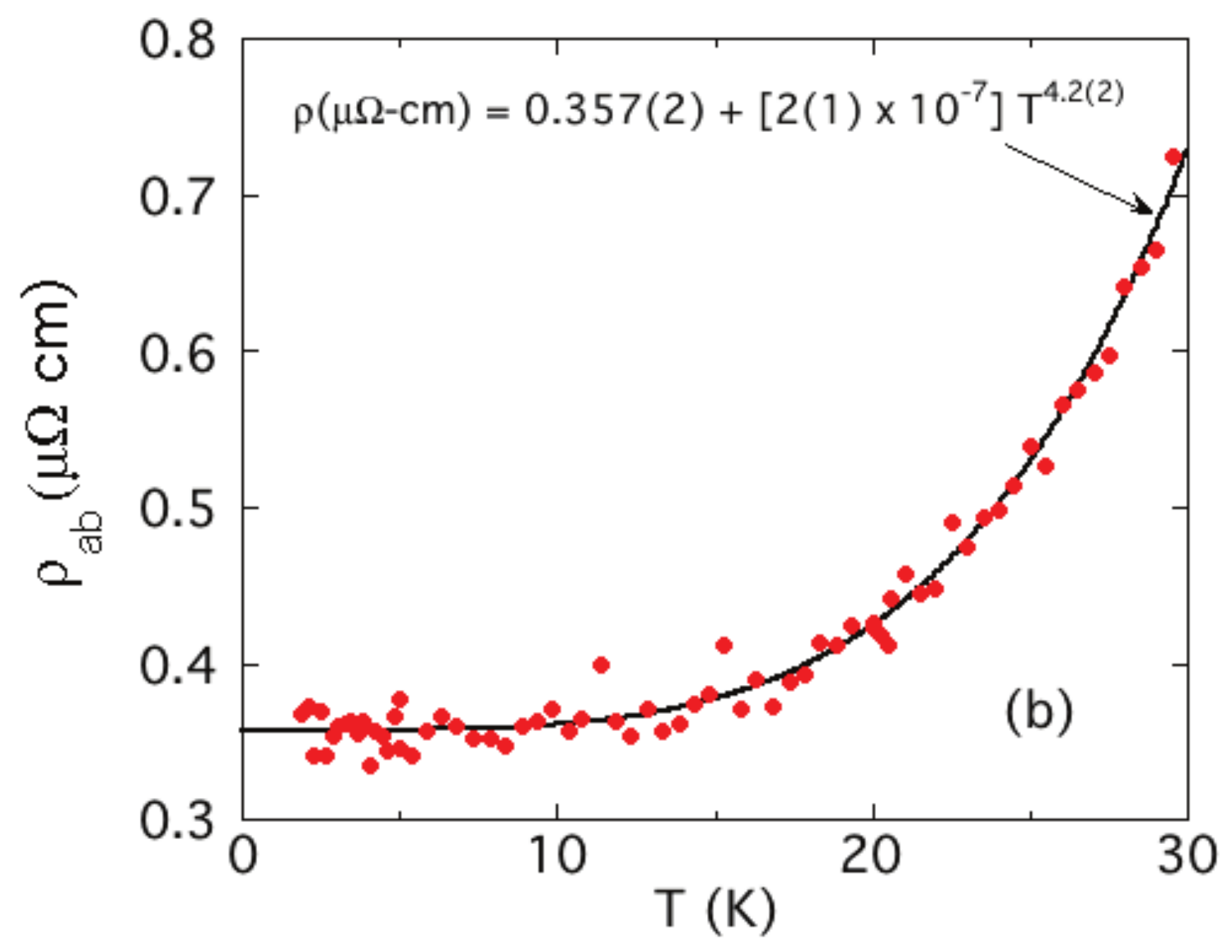}
\caption{(a) Basal $ab$-plane electrical resistivity $\rho_{ab}$ of KCo$_2$As$_2$ versus temperature $T$ from 2 to 300~K (red data).  Also shown is the best fit of the data by the Bloch-Gr\"uneisen (BG) model (black data). The residual-resistivity ratio (RRR) is $\rho$(300~K)/$\rho(T\to0)$. The $\rho(T\to0)$ was obtained from panel~(b). The large RRR indicates high purity of the crystal measured.  (b)~Expanded plot of the data below 30~K (filled red circles), together with the fit by Eq.~(\ref{Eq:rhoLowT}).  The data below 30~K approximately follow a $T^4$ dependence as shown, instead of the $T^5$ dependence expected from the BG model. }
\label{Fig:RvsTfits}
\end{figure}

\begin{figure}
\includegraphics[width=3in]{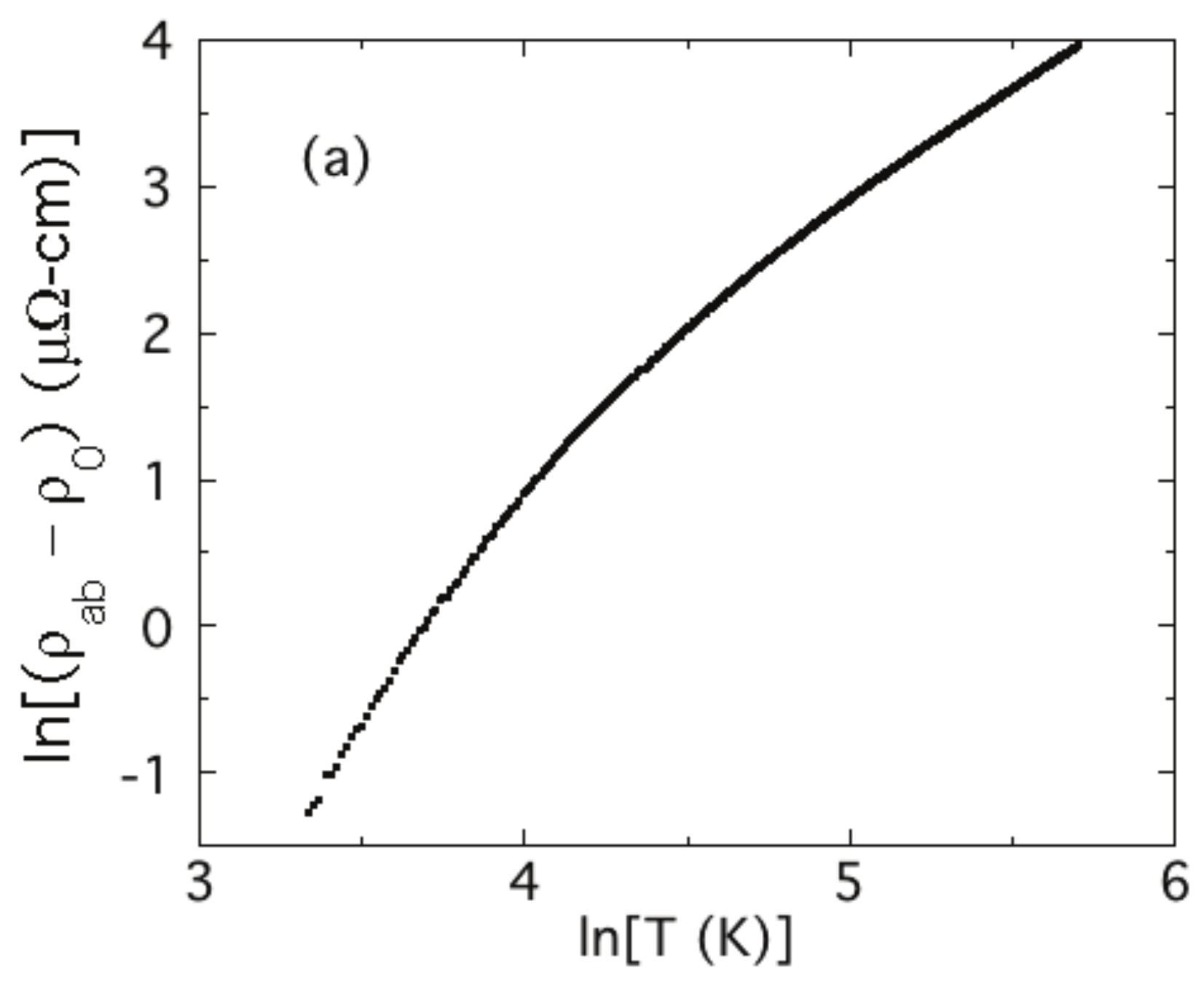}
\includegraphics[width=3in]{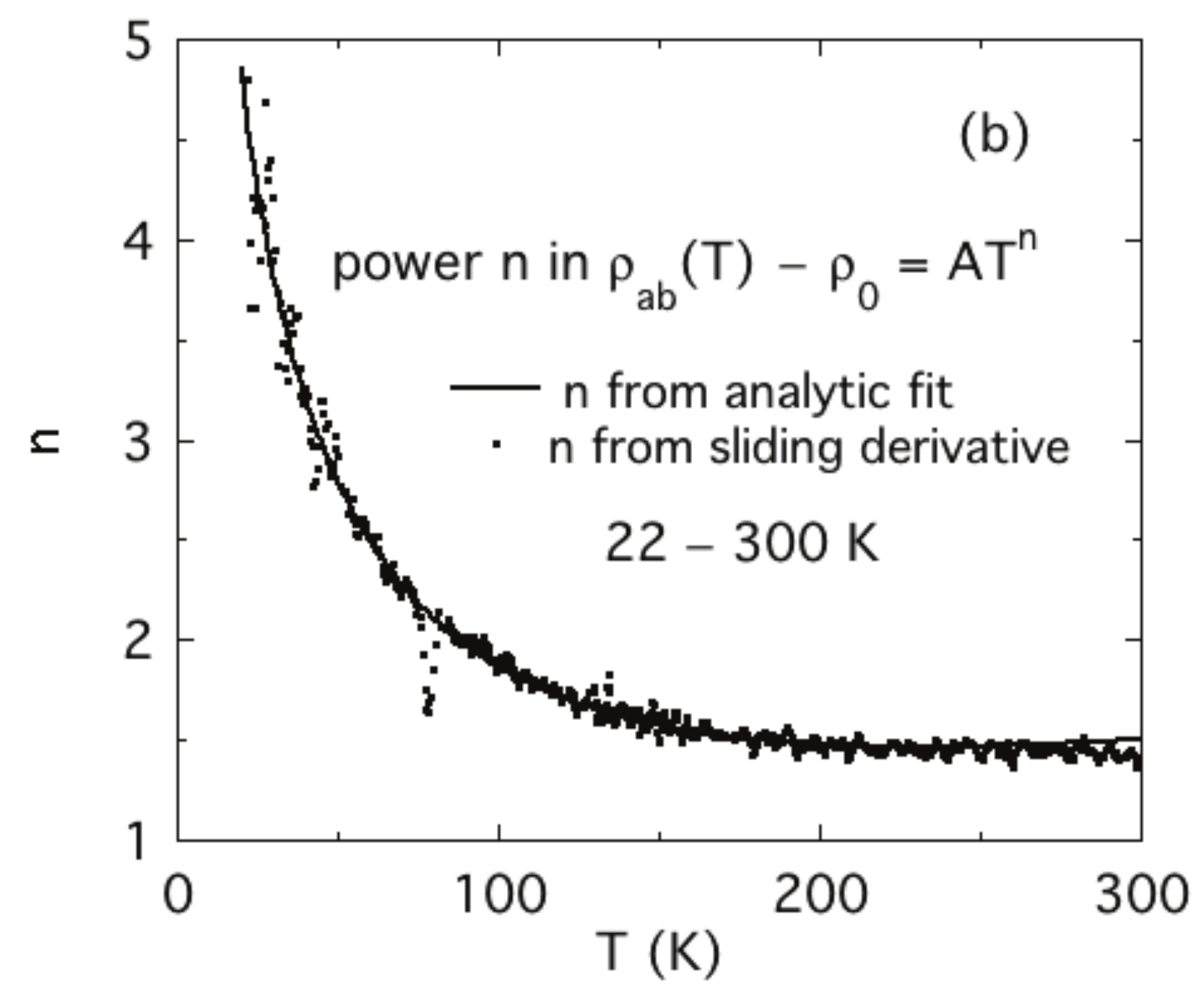}
\caption{(a) Logarithm of $\rho_{ab}(T)-\rho_{0}$ versus the logarithm of $T$ from 2 to 300~K (red data) [$\rho_{0}$ is the residual resistivity at \mbox{$T=0$~K} from Fig.~\ref{Fig:RvsTfits}(b)].  The nonlinear behavior shows that the temperature dependence $T^n$ does not have a $T$-independent exponent~$n$ over this $T$ range.  (b)~The $T$ dependence of $n$ obtained from the data in~(a) using Eq.~(\ref{Eq:n(T)}) and also a sliding derivative. The value $n\sim 4$ at 22~K, averaging over the scatter in the sliding derivative, agrees with the value of the exponent in the equation in Fig.~\ref{Fig:RvsTfits}(b); the source of the clear anomaly at 78~K in these data is unclear as yet.}
\label{Fig:nVsT}
\end{figure}

The basal $ab$-plane resistivity $\rho_{ab}$ of \kca\ over the full $T$ range of the measurements 2--300~K is shown in Fig.~\ref{Fig:RvsTfits}(a).  Also shown is a fit of the data by the Bloch-Gr\"uneisen (BG) model for the resistivity arising from electron scattering from acoustic phonons using the high-accuracy Pad\'e-approximant fit to the BG model given in Ref.~\cite{Goetsch2012}.  It is evident that the BG model does not describe the high-$T$ behavior of the $ab$-plane resistivity.  In particular, the BG model predicts a linear $\rho(T)$ at high temperatures.  Instead the data show positive curvature from 30~K to 300~K\@.

As shown in Fig.~\ref{Fig:RvsTfits}(b), the data below 30 K are fitted by the expression
\bea
\rho_{ab}(T) = \rho_0 + A T^n,
\label{Eq:rhoLowT}
\eea
where the fitted values of residual resistivity $\rho_{0}$, $A$ and $n$ are given in the figure with error bars.  The low-$T$ temperature exponent is $n= 4.2(2)$ instead of the value of 5  predicted by the BG model.  Interestingly, the exponent of $n=4$ is the value expected for two-dimensional electron-phonon scattering at low $T$ as in graphene~\cite{Park2014}.  The $ab$-plane residual resistivity is 0.357\,$\mu\Omega$\,cm for $T\to0$, which is a factor of 34 smaller than that in the isostructural SrCo$_2$As$_2$~\cite{Pandey-2013b} and a factor of 14 smaller than that in BaCo$_2$As$_2$~\cite{Anand-2014a}.

The natural log of $\rho_{ab}(T)-\rho_{ab0}$ of KCo$_2$As$_2$ is plotted versus the natural log of $T$(K) in Fig.~\ref{Fig:nVsT}(a).  The nonlinear behavior shows that a single exponent~$n$ in $T^n$ cannot describe the $T$ dependence.  Therefore we analyzed the data in Fig.~\ref{Fig:nVsT}(a) as follows.  Writing $\rho_{ab}(T)-\rho_{0}=AT^n$ where $A$ is a constant, one obtains
\bea
n(T) = \frac{d \ln[\rho_{ab}(T)-\rho_{0}]}{d \ln T},
\label{Eq:n(T)}
\eea
where $n(T)$ is thus the slope of the plot in Fig.~\ref{Fig:RvsTfits}(a) at each $T$\@.  First, we fitted $\ln(\rho_{ab}-\rho_0)$ vs~$\ln T$ in Fig.~\ref{Fig:nVsT}(a) by a third-order polynomial and obtained $n(T)$ by differentiating that function as shown by the solid curve in Fig.~\ref{Fig:nVsT}(b).  Alternatively, we carried out a pointwise sliding derivative of the $\ln(\rho_{ab}-\rho_0)$ vs~$\ln T$ where the derivative was calculated as the slope in the region from $T-\Delta T$ to $T+\Delta T$, where $\Delta T \approx 2$~K\@.  The result is shown as the filled circles in Fig.~\ref{Fig:nVsT}(b).  There is a dip in the latter data at about 78~K, close to the boiling point (77.3~K) of liquid N$_2$ at 1~atm pressure.  Additional measurements are required to determine whether or not this feature in Fig.~\ref{Fig:nVsT}(b) is intrinsic to \kca.  In the sliding derivative calculation, $n$ is seen to decrease from $\sim 4$ at 22~K, consistent with the power $n=4.2(2)$ in the equation in Fig.~\ref{Fig:RvsTfits}(b), to a value of about 1.4 at 300~K where $\rho_{ab}(T)$ therefore still exhibits positive curvature.  

\begin{figure}
\includegraphics[width=3in]{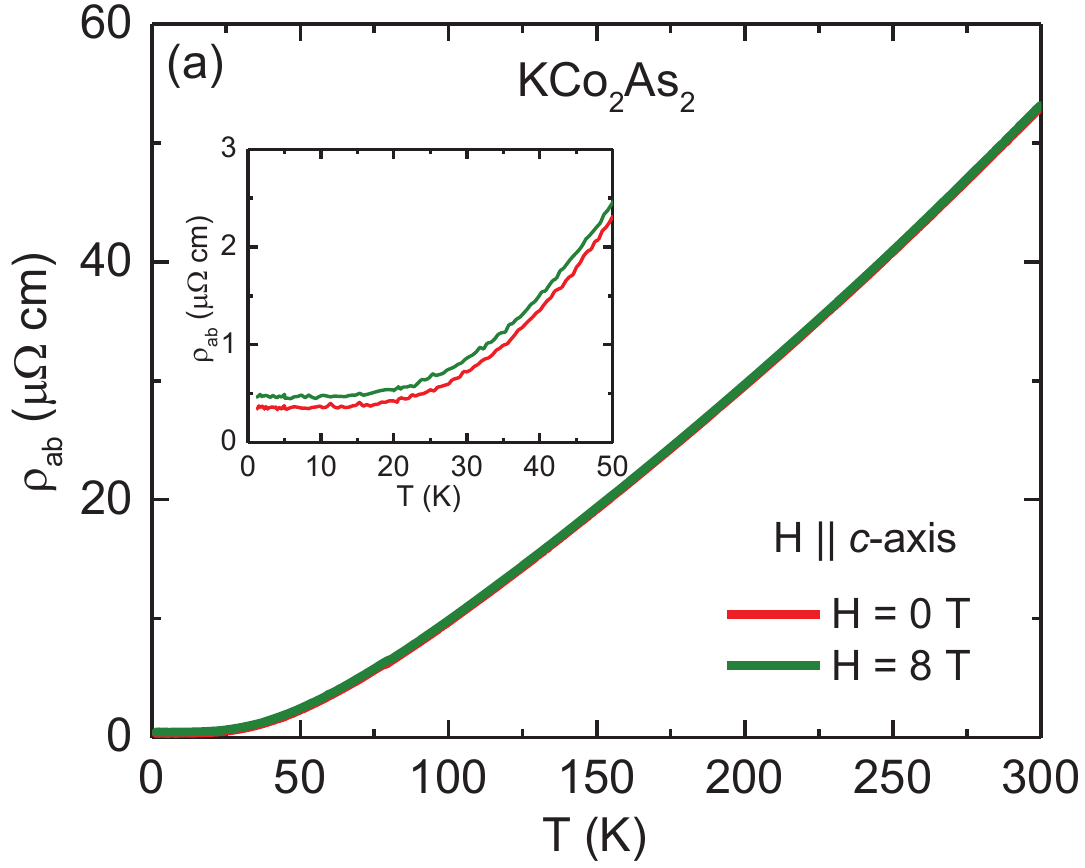}
\includegraphics[width=3in]{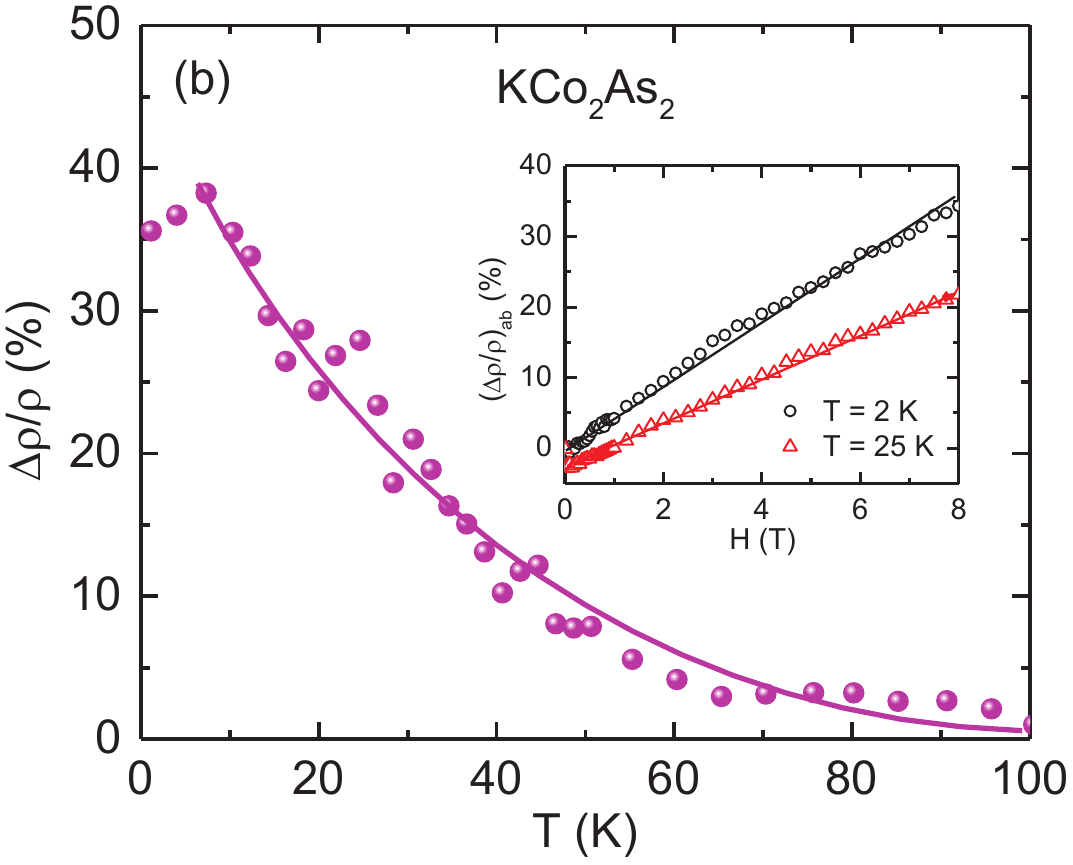}
\caption{(a) Basal $ab$-plane electrical resistivity $\rho_{ab}$ versus temperature $T$ of KCo$_2$As$_2$ measured in two different magnetic fields $H = 0$ and 8~T applied along the crystallographic $c$-axis. Inset: The $\rho_{ab}(T)$ at $H = 0$ and 8~T for $T \le 50$~K\@. (b)~Magnetoresistance $(\Delta\rho/\rho)_{ab}$ versus $T$, where $\Delta\rho_{ab}$ is the change in resistivity induced by an external 8~T magnetic field. Inset: $(\Delta\rho/\rho)_{ab}$  versus $H$ measured at two different temperatures, $T = 2$ and 25~K\@. The solid lines are guides to the eye.}
\label{fig:MR}
\end{figure}

A comparison of the $\rho_{ab}(T)$ data taken at two different external magnetic fields $H = 0$ and 8~T is shown in Fig.~\ref{fig:MR}(a). While at higher temperatures ($T \gtrsim 100$~K) the two data plots nearly overlap each other, they show a sizable deviation at low temperatures where $\rho_{ab,\rm 8\,T} > \rho_{ab,\rm 0\,T} $ [inset, Fig.~\ref{fig:MR}(a)]. The magnetoresistance $(\Delta\rho/\rho)_{ab}$ versus $T$ is shown in Fig.~\ref{fig:MR}(b). The $(\Delta\rho/\rho)_{ab}$ exhibits a value of about $35\%$ at 2~K, then shows a modest increase with increase in $T$, then monotonically decreases with increasing $T$ before becoming negligible at $\gtrsim100$~K\@. Additionally, the $(\Delta\rho/\rho)_{ab}$ varies nearly linearly with $H$ at low temperature [inset, Fig.~\ref{fig:MR}(b)]. The observed large value of $(\Delta\rho/\rho)_{ab}$ at 2~K and $H=8$~T and its nearly linear variation with $H$ warrants additional measurements of this quantity at higher fields.

\subsection{\label{Cp} Heat capacity}

\begin{figure}
\includegraphics[width=3.in]{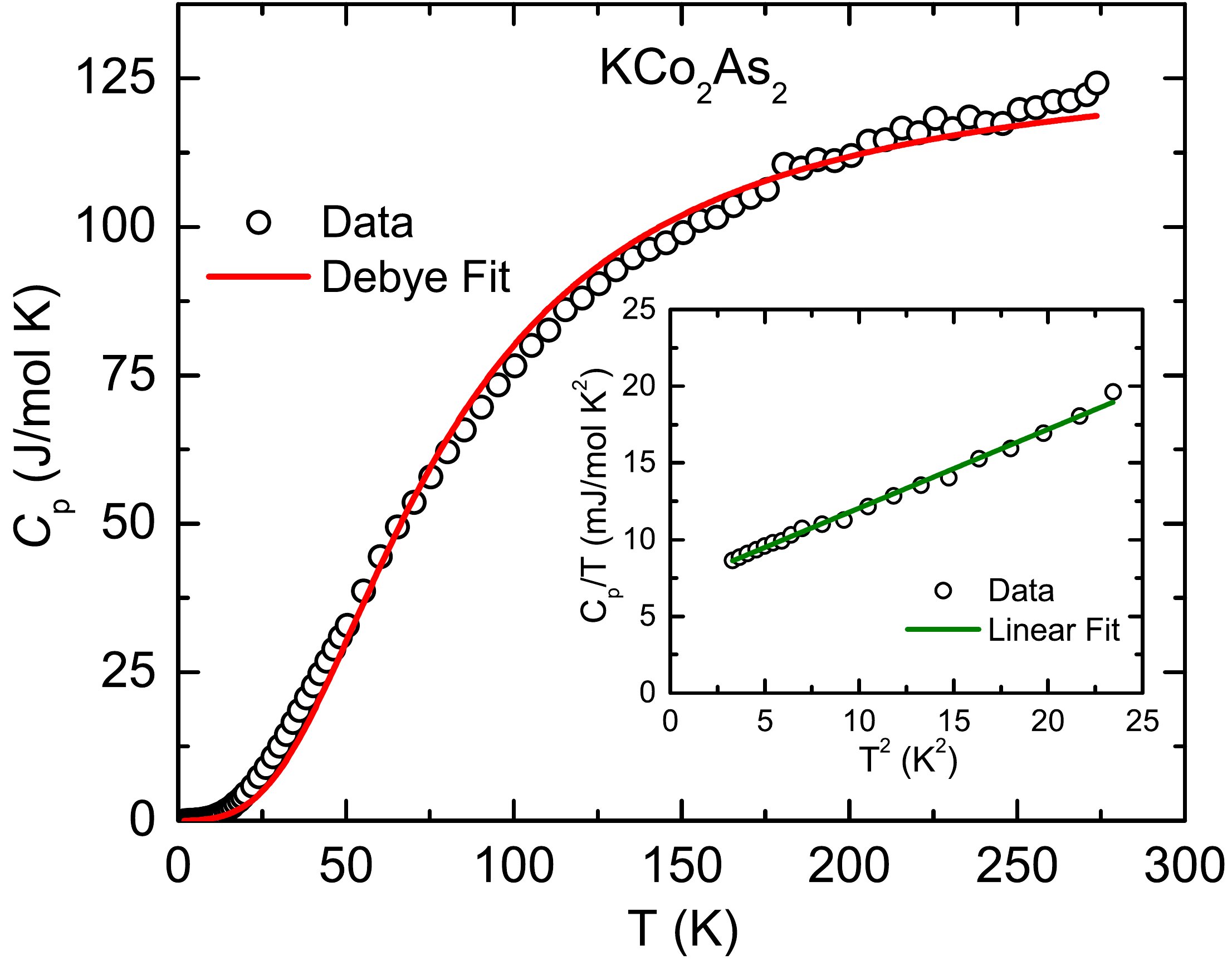}
\caption{Heat capacity $C_{\rm p}$ versus temperature $T$ of a single crystal of KCo$_2$As$_2$. The red curve is a fit of the data by the Debye model. Inset: plot of $C_{\rm p}(T)/T$ versus $T^2$. The green line is a linear fit to the data. The fit is poor.  The fact that the data do not asymptote to a constant value by 300~K suggests that a heat capacity component from optic modes may be  present.}
\label{fig:HC}
\end{figure}

The heat capacity $C_{\rm p}$ versus $T$ of KCo$_2$As$_2$ in Fig.~\ref{fig:HC} does not show evidence of any phase transition and exhibits a value of 124~J/mol\,K$^2$ at 274~K\@. This value is close to the Dulong-Petit high-$T$ limit of the lattice heat capacity at constant volume given by $C_{\rm V} = 3nR = 124.7$~J/mol\,K$^2$, where $n = 5$ is the number of atoms per formula unit  and $R$ is the molar gas constant. The data were fitted by the Debye model for acoustic phonons \cite{Goetsch2012,Gopal-1966} as shown by the solid curve in Fig.~\ref{fig:HC}. The fitted value of the Debye temperature $\Theta_{\rm D}$ is 316(2)~K (Table~\ref{Table:Parameters}). However, it is evident from Fig.~\ref{fig:HC} that the fit is not optimum, indicating that the Debye model does not adequately describe the phonon spectrum.  We suggest that excitation of optic vibration modes may occur.

At low temperatures, the heat capacity can be described by $C_{\rm p}(T) = \gamma T  + \beta T^3$, where $\gamma$ is the electronic  Sommerfeld coefficient and $\beta$ is the coefficient associated with the lattice contribution. The $C_{\rm p}(T)/T$ versus $T^2$ plot for KCo$_2$As$_2$ in the inset of Fig.~\ref{fig:HC} shows a linear behavior for $T \le 5$~K and the fit using the above expression (green line in the inset) yields $\gamma = 6.9(1)$~mJ/mol\,K$^2$ and $\beta = 0.514(9)$~mJ/mol\,K$^4$. The value of $\gamma$ is significantly smaller than those for SrCo$_2$As$_2$ and BaCo$_2$As$_2$ (Table~\ref{Table:Parameters}). The low-$T$ $\Theta_{\rm D}$ calculated from the value of $\beta$ using $\Theta_{\rm D} = (12\pi^4Rn/5\beta)^{1/3}$ is 266(2)~K, where $R$ is the molar gas constant and $n = 5$ is the number of atoms per formula unit. 

The electronic density of states at Fermi level for both spin directions including many-body enhancement effects due to electron-electron and/or electron-phonon interactions ${\cal D}_\gamma(E_{\rm F})$ can be estimated from $\gamma$ using
\begin{equation}
{\cal D}_\gamma(E_{\rm F}) = \frac{3\gamma}{\pi^2k_{\rm B}^2},
\label{eq:2}
\end{equation}
where $k_{\rm B}$ is Boltzmann's constant. If $\gamma$ is expressed in units of mJ/mol\,K$^2$, then ${\cal D}_\gamma(E_{\rm F})$ can be written 
\begin{equation}
{\cal D}_\gamma(E_{\rm F}) = \frac{\gamma}{2.359}~\frac{\rm states}{\rm eV\,f.u.}.
\label{eq:3}
\end{equation}
The value of ${\cal D}_\gamma(E_{\rm F})$ of KCo$_2$As$_2$ calculated using Eq.~(\ref{eq:3})
and $\gamma = 6.9(1)$~mJ/mol\,K$^2$ is 2.92(5)~states/(eV\,f.u.).
This is significantly smaller than those reported for other $A$Co$_2$As$_2$ ($A =$ Ca, Sr, Ba) compounds
(Table~\ref{Table:Parameters}),
but is significantly larger than the value ${\cal D}(E_{\rm F})= 2.15$~states/(eV\,f.u.)\@ 
obtained from our band-structure calculations for \kca\ in Sec.~\ref{Theory} below.
The specific heat enhancement relative to the bare band structure density of states value can
be written as

\begin{equation}
(1+\lambda_\gamma)m^\ast/m_{\rm band} = \frac{\gamma}{\gamma_{\rm band}} = \frac{{\cal D}_\gamma(E_{\rm F})}{{\cal D}_{\rm band}(E_{\rm F})},
\label{eq:4}
\end{equation}
where $\lambda_\gamma$ is a coupling constant that usually arises from electron-phonon coupling but can also have contributions from other degrees of freedom, specifically spin~fluctuations in some materials \cite{brinkman,mazin},
$m^\ast/m_{\rm band}$ is a remaining band renormalization not from coupling to low energy bosons, and $\gamma_{\rm band}$
is the bare band structure value of the Sommerfeld coefficient. The calculated value of $(1+\lambda_\gamma)m^\ast/m_{\rm band}$ using Eq.~(\ref{eq:4}) is 1.36(3).  This value indicates a substantial enhancement from free-electron behavior and is comparable to the value of 1.45 for ${\rm SrCo_2As_2}$ in Table~\ref{Table:Parameters}.

\subsection{\label{Magnetism} Magnetic properties}

\begin{figure}
\includegraphics[width=3.in]{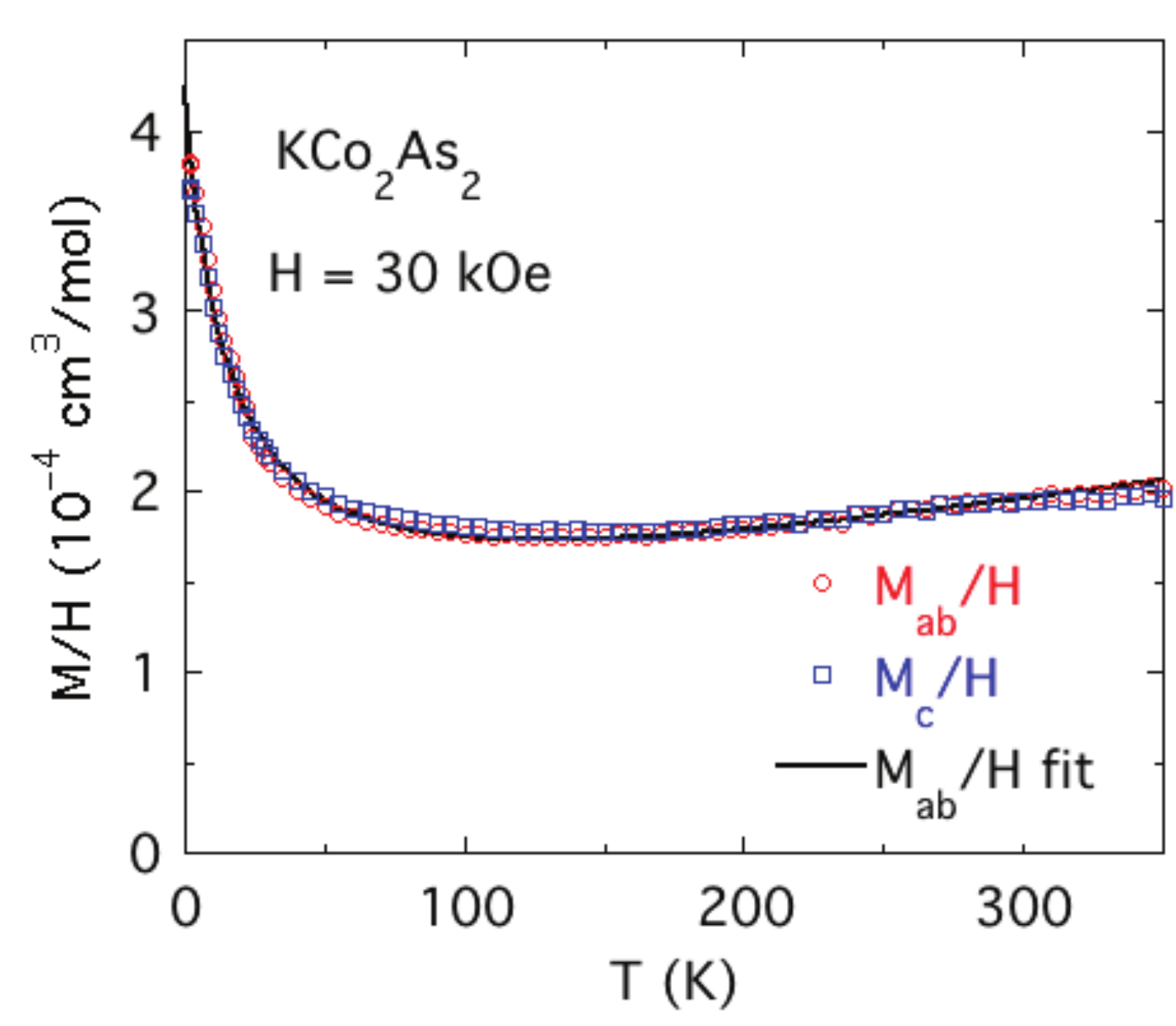}
\caption{Magnetization divided by field $M/H$ versus temperature $T$ of a crystal of KCo$_2$As$_2$ with the applied field $H=30$~kOe aligned in the $ab$~plane (open red circles) and along the $c$~axis (open blue squares).  The data are seen to be nearly isotropic.  A very good fit to the $M_{ab}/H$ vs~$T$ data is obtained using Eq.~(\ref{Eq:ChiFit}) and is shown as the black curve.  The parameters of the fit are given in the text.}
\label{fig:chi_H_3_T}
\end{figure}

The magnetic susceptibilities $M/H$ versus~$T$ in an applied field  $H=30$~kOe applied along the $ab$~plane and along the $c$~axis are plotted in Fig.~\ref{fig:chi_H_3_T}.  The data are seen to be nearly isotropic.  Above 100~K, the data show an increase with increasing~$T$ which is  likely associated with an increase in the density of states at the Fermi level with increasing~$T$ as discussed below in Sec.~\ref{Theory}, whereas below about 100~K a Curie-Weiss-like upturn occurs which we assume is associated with magnetic impurities or defects in the crystal. Therefore the $ab$-plane data were fitted over whole $T$ range by
\bea
M_{ab}(T)/H = \chi_0 + A T + M_{\rm imp}(T)/H,
\label{Eq:ChiFit}
\eea
where $\chi_0$ is the intrinsic  contribution at $T=0$, $AT$ is the above-noted weak positive contribution linear in $T$ where $A = 7.3(3)\times10^{-7}$\,cm$^3$/(mol\,K).  The term  $M_{\rm imp}(T)/H$ is the contribution of the magnetic impurities, where we used a Brillouin function for assumed spin-1/2 impurities/defects with $g=2$ to determine the contribution of this term.  The fitted parameters of the Brillouin function were the prefactor $f=1.33(6)$\% which is the molar fraction of the $S=1/2$ impurity/defect spins with respect to KCo$_2$As$_2$.  We also replaced $T$ in the Brillouin function by $T-\theta_{\rm imp}$ to treat possible interactions between the impurity spins in a simple way.  We obtained the Weiss temperature $\theta_{\rm imp} = -15.7(7)$~K which corresponds to antiferromagnetic interactions between the local-moment impurities/defects.

\begin{figure}
\includegraphics[width=3in]{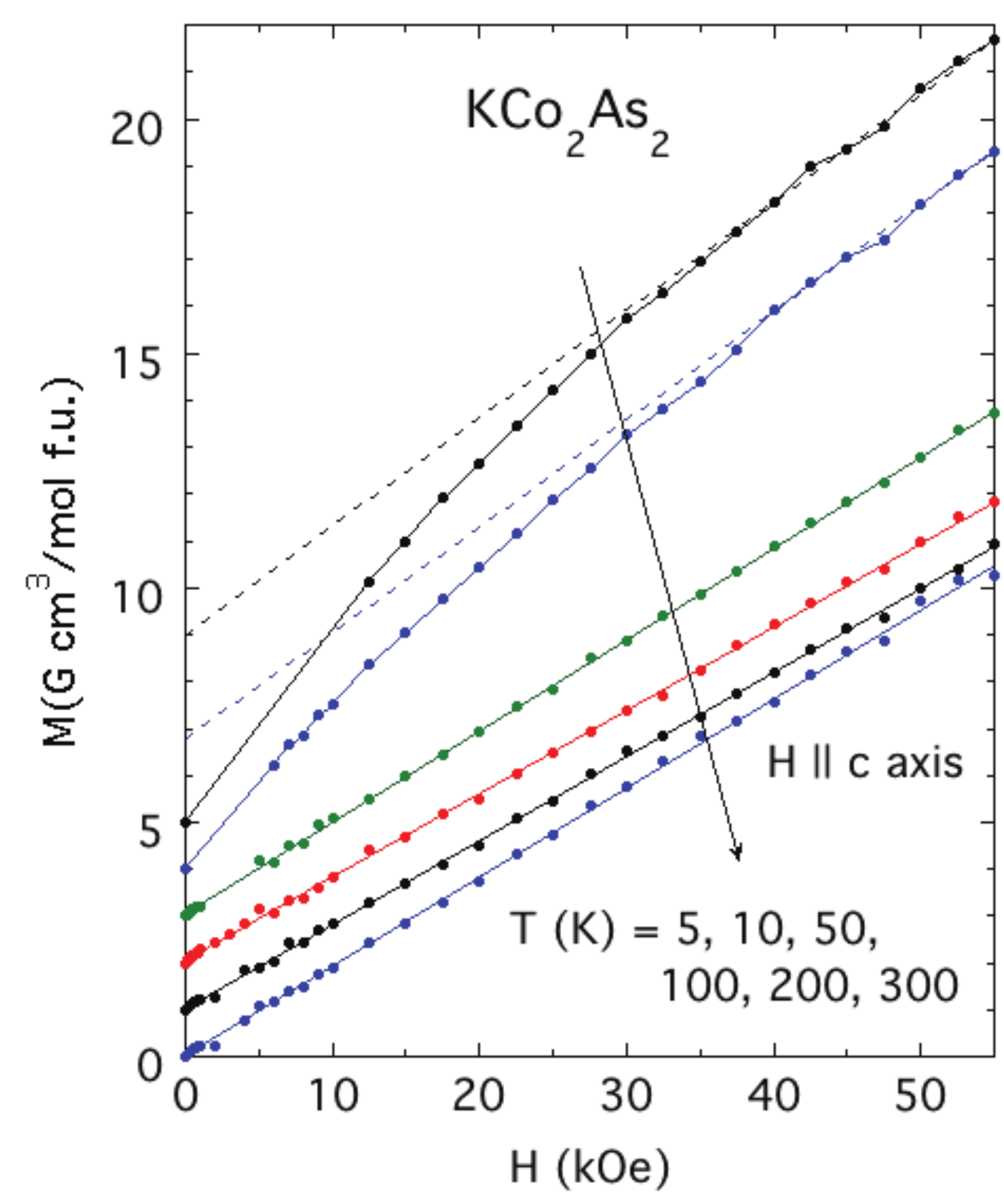}
\caption{Isothermal magnetization $M$ versus $H$ ($\parallel c$-axis) measured at six different temperatures.  The data with decreasing temperatures are offset successively upwards by 1~G\,cm$^3$/mol for clarity.  The linear fits for $T=50$--300~K and $H = 0$ to 55~KOe are shown as the straight lines, whereas the linear fits to the data for 5 and 10~K between 40 and 55~kOe are shown as the dashed lines with extrapolations to $H=0$. The linear fit parameters are given in Table~\ref{Table:M(H)}.}
\label{fig:Mag}
\end{figure}

\begin{table}
\caption{Linear-fit parameters for the $M(H)$ isotherm data for KCo$_2$As$_2$ according to Eq.~(\ref{Eq:M(H)fit}).  Also included is the density of states at the Fermi level $N_\chi(E_{\rm F})$ for both spin directions for $T=100$--300~K according to Eq.~(\ref{eq:7}), and the bare band structure value from Fig.~\ref{Fig:Fermi_Level_vs_T} below.}
\label{Table:M(H)}
\begin{ruledtabular}
\begin{tabular}{ccccc} 
$T$ & $H$\,Fit\,Range & $M(H=0)$  & $\chi$ & $N_\chi(E_{\rm F})$ \\
(K)  & (kOe) & (G\,cm$^3$/mol) & ($10^{-4}$\,cm$^3$/mol) & ${\rm \frac{states}{eV\,f.u.}}$ \\
\hline
5 & 40--55 & 4.0(4) & 2.311(9) \\
10 & 40--55 & 2.8(3) & 2.282(7) \\
50 & 0--55 & 0.05(2) & 1.951(6) \\
100 & 0--55 & 0.05(2) & 1.781(6) & 5.51, 2.15\\
200 & 0--55 & 0.04(2) & 1.790(6) & 5.54, 2.17\\
300 & 0--55 & 0.04(2) & 1.902(8) & 5.88, 2.22\\
\end{tabular}
\end{ruledtabular}
\end{table}

Isothermal magnetization $M$ versus $H$ measurements at six temperatures below 300~K are plotted in Fig.~\ref{fig:Mag}. The $M$ is proportional to $H$ for $T \geq 50$~K\@. However, we observe negative curvature in the 5~K and 10~K  data. The upturn in the $\chi(T)$ data and the negative curvature at low temperatures in the $M(H)$ data suggest the presence of paramagnetic impurities or defects. Such magnetic impurities are expected to saturate at high fields. Hence, to determine the intrinsic $\chi$ at low temperatures, we calculated the slopes of the linear $M(H)$ data in a $T$-dependent field range according to
\bea
M(T, H) = M(T, H=0) + \chi(T)H,
\label{Eq:M(H)fit}
\eea
where $M$ is in cgs units of G\,cm$^3$/mol\,f.u., $\chi$ is in units of cm$^3$/mol, and $H$ is in units of Oe. The Pauli formula gives a bare susceptibility in terms of the density of states of a non-interacting electron gas. The corresponding
density of states at $E_{\rm F}$ for both spin directions ${\cal D}_\chi(E_{\rm F})$ can be obtained from $\chi$ and the Pauli formula
\begin{equation}
N_\chi(E_{\rm F}) = \frac{\chi}{\mu_{\rm B}^2},
\label{eq:6}
\end{equation}
where $\mu_{\rm B}$ is Bohr magneton, $\chi$ is expressed in units of cm$^3$/mol, and a spectroscopic splitting factor $g=2$ of the conduction-carrier spins is assumed.
We obtain
\begin{equation}
N_\chi(E_{\rm F}) ({\rm states/eV\,f.u.})= 3.093\times10^4\,\chi\,(\rm cm^3/mol)
\label{eq:7}
\end{equation}
for both spin directions. These values are listed as the first entries in the last column of Table~\ref{Table:M(H)}. The bare band-structure densities of states at 100, 200, and 300~K from Fig.~\ref{Fig:Fermi_Level_vs_T} below are the second entries in the last column of Table~\ref{Table:M(H)}.  We see that both the experimental and theoretical $N(E_{\rm F})$ values increase with increasing~$T$, with the experimental values about 2.6 times larger than the bare theoretical values.  This feature will be discussed further in the following section.  The listed values of the Pauli susceptibility $\chi$ in Table~\ref{Table:M(H)} are likely a bit too small, because we have not corrected for the presence of diamagnetism of the atomic cores and of the conduction carriers.  We also note that the free-electron model in two and three dimensions predicts that $\chi$ decreases with increasing~$T$~\cite{Johnston2020} instead of increasing as found here.

Inserting the value of $\chi$(200~K) of KCo$_2$As$_2$ from Table~\ref{Table:M(H)} into Eq.~(\ref{eq:7}) yields ${\cal D}_\chi(E_{\rm F}) = 5.54$~states/eV\,f.u. The Wilson ratio $R_{\rm W} = {\cal D}_\chi(E_{\rm F})/{\cal D}_\gamma(E_{\rm F})$ is then estimated to be 1.88.  This is significantly smaller than the values of 3.4 and 5.2 for SrCo$_2$As$_2$ and BaCo$_2$As$_2$, respectively (Table~\ref{Table:Parameters} below).

\subsection{\label{Theory} Theory}

For an itinerant-electron system, the bare Pauli susceptibility is renormalized according to the
Stoner formula to yield a Wilson ratio $R_{\rm W}$ different from unity. The Stoner formula for $R_{\rm W}$ is
\begin{equation}
R_{\rm W} = \frac{1}{1 - N(E_{\rm F})I},
\end{equation}
where $N(E_{\rm F})$ is the density of states at the Fermi level on a per transition metal atom, per spin direction basis,
and $I$ is an interaction parameter that is typically $\sim1$~eV or less for $3d$ transition metals~\cite{janak}.
Inserting the calculated first-principles density of states described below and taking $I= 1$~eV yields $R_{\rm W}= 2.16$, in reasonable accord with the experimental value of 1.88 in Table~\ref{Table:Parameters} below. However, this calculated value may be overestimated due to the presence of hybridization between Co and As states, which would reduce the effective Stoner~$I$.

\begin{figure}[h]
\includegraphics[width=3.3in]{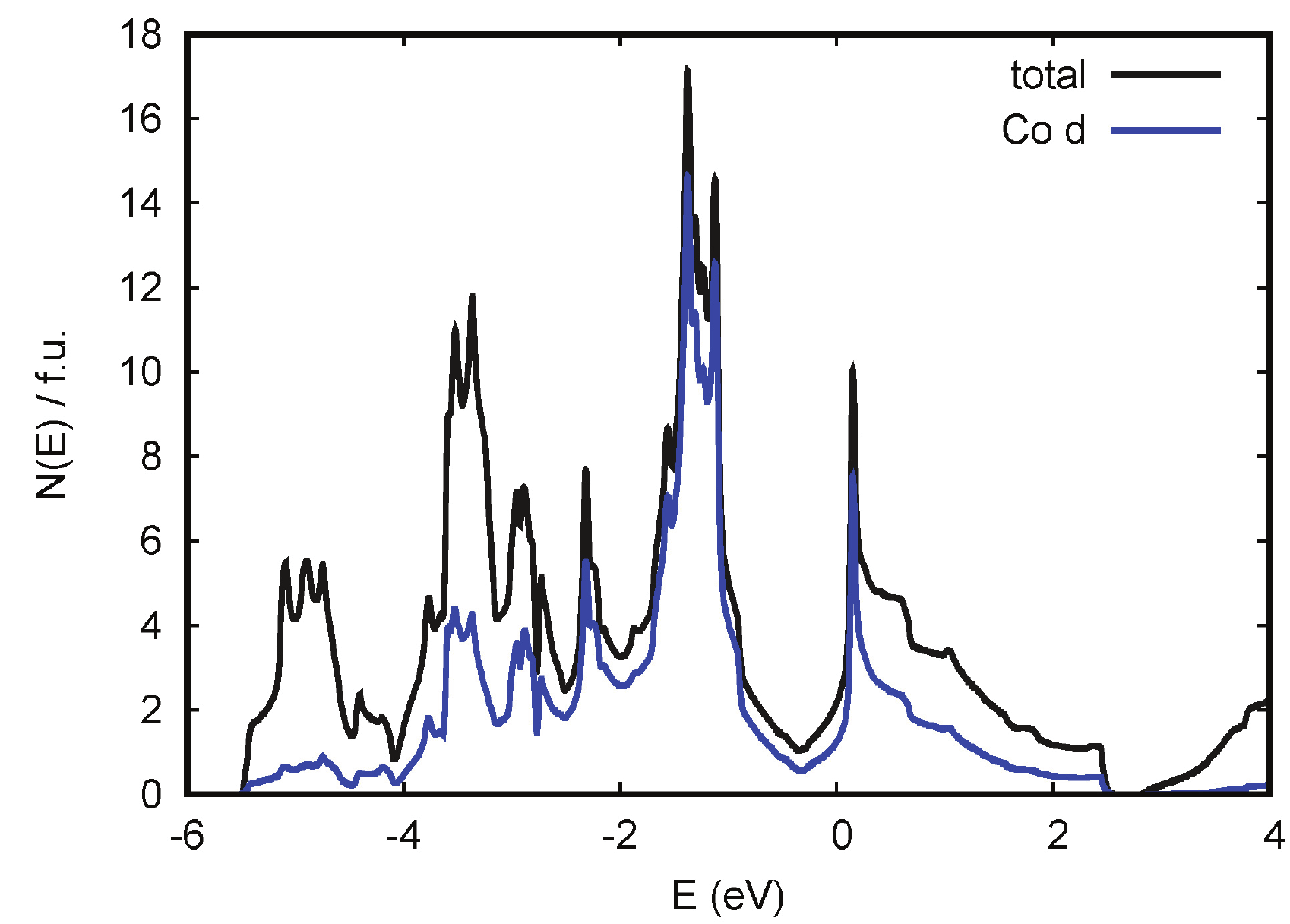}
\caption{Total density of states $N(E)$/f.u.\ versus energy $E$ (black line) in units of states/eV\,f.u.\ for both spin directions.  The zero of energy is the Fermi energy. Also shown is the contribution from the Co $3d$~states (blue line).}
\label{Fig:DOS}
\end{figure}

The calculated density of states versus energy is shown in Fig.~\ref{Fig:DOS} in units of states/eV-f.u.\ for both spin directions.  Also shown is the contribution from the Co~$d$~states.  These calculations are more accurate than our previous ones for the same compound in Ref.~\cite{Singh-2009}, since we now include spin-orbit coupling.  The density of states at the Fermi energy is $N(E_{\rm F})=2.15$~states/(eV\,f.u.).

\begin{figure}[t]
\includegraphics[width=2.5in]{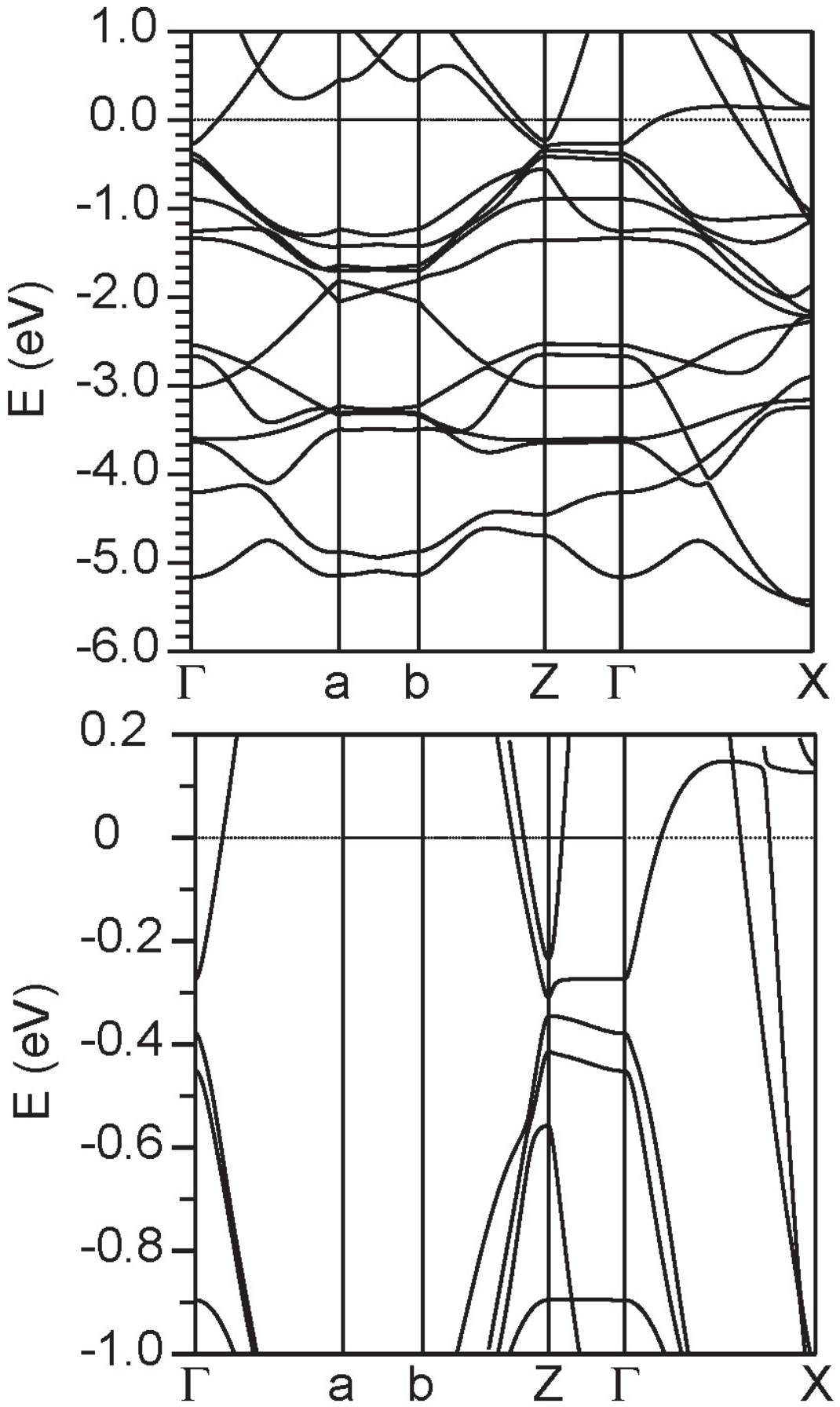}
\caption{Band Structure of \kca.  The points in the Brillouin zone are defined in Fig.~\ref{Fig:BrillouinZone}.  The top panel shows an overall view and the bottom panel shows an expanded plot below the Fermi energy.}
\label{Fig:BandStruct}
\end{figure}

\begin{figure}
\includegraphics[width=3.in]{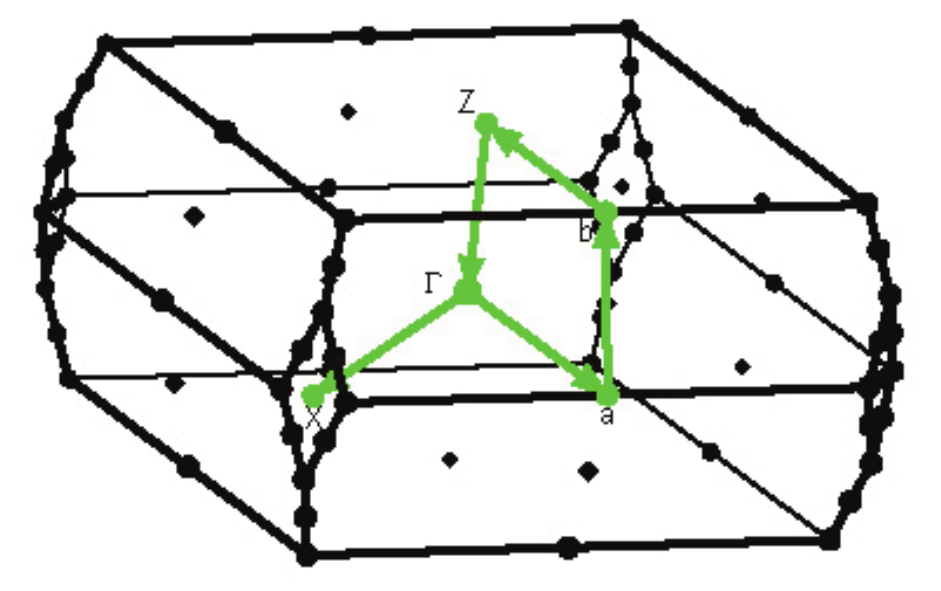}
\caption{Brillouin zone for the bct structure of \kca\ with points in the Brillouin zone marked.}
\label{Fig:BrillouinZone}
\end{figure}

The band structure is shown in the top panel of Fig.~\ref{Fig:BandStruct} and an expanded plot below the Fermi energy $E_{\rm F}$ is shown in the bottom panel.  The notation for points in the bct Brillouin zone (BZ) is indicated in Fig.~\ref{Fig:BrillouinZone}. Of particular note is the flat band between $\Gamma$ and~X which gives rise to the sharp peak  in the density of states versus energy at about 0.1~eV above $E_{\rm F}$ in Fig.~\ref{Fig:DOS}.  Two bands cross $E_{\rm F}$ near the Z and $\Gamma$ points of the BZ\@. The occupations of these two bands are respectively 0.316 and 0.184 electrons per spin direction. This leads to a total of 0.5 electrons per spin direction, i.e. 1 electron for both spin directions, corresponding to the electron contributed by the K$^{1+}$ ion.

\begin{figure}
\includegraphics[width=2.25in]{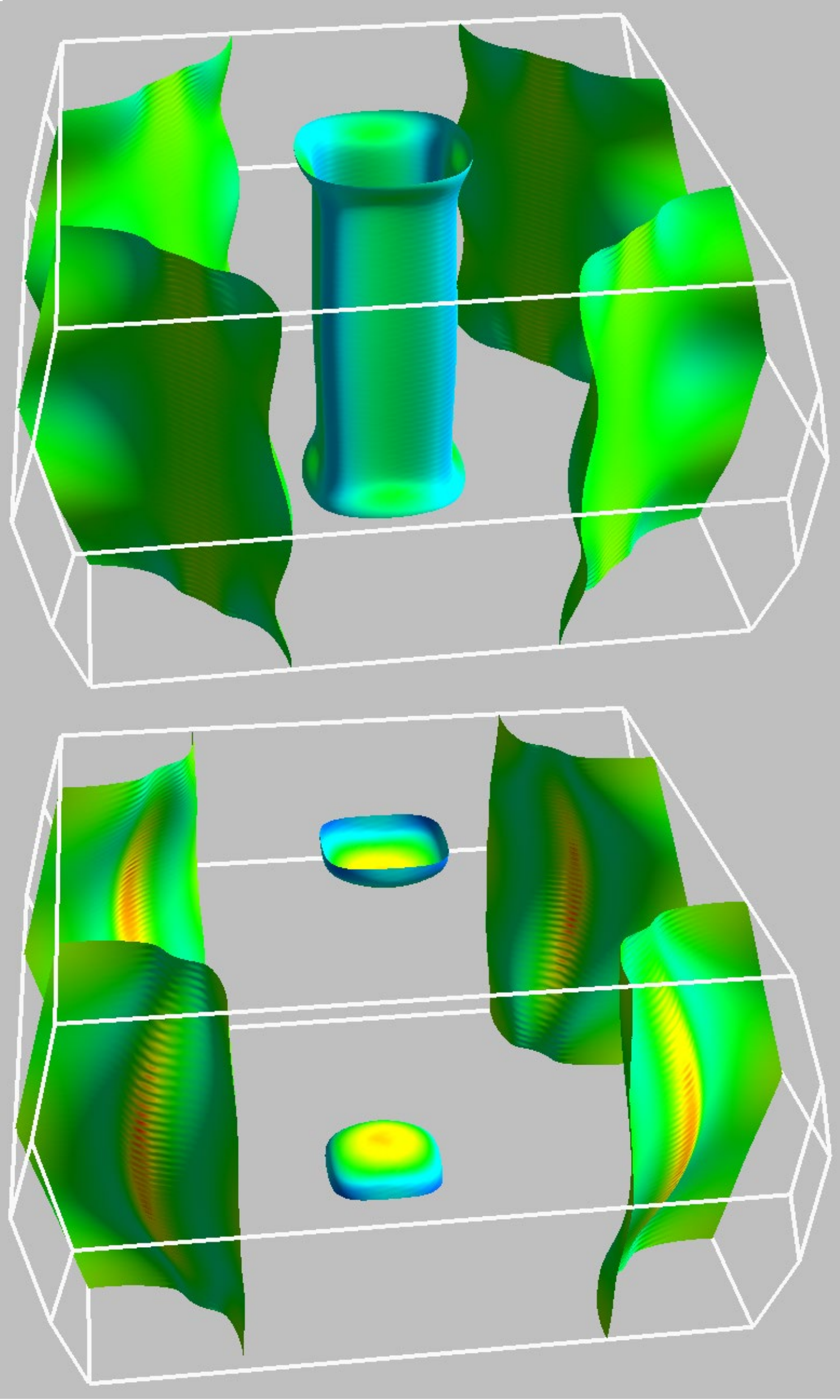}
\caption{Fermi surfaces of \kca, shown in perspective view, with $\Gamma$ at the center and the zone edges indicated
by white lines. The lower band is in the top panel and the upper band is in the lower panel.  The shading is by velocity, $\nabla_{\bf k}E({\bf k})/\hbar$, where blue is lower velocity and green is higher velocity. The Fermi surfaces are electron-like. The lower band has an occupancy of 0.316 of the zone, while the upper band has an occupancy of 0.184 of the zone.}
\label{Fig:Fermi_Surfaces}
\end{figure}

The Fermi surfaces for the lower and upper bands are shown in Fig.~\ref{Fig:Fermi_Surfaces}. High-velocity electron barrels occur along the zone corners. These dominate the in-plane conduction. A lower-velocity electron cylinder occurs at the zone center in the lower band. The upper band has a three-dimensional section with higher velocity around the Z points. This is important for $c$-axis conduction.

\begin{figure}
\includegraphics[width=3.3in]{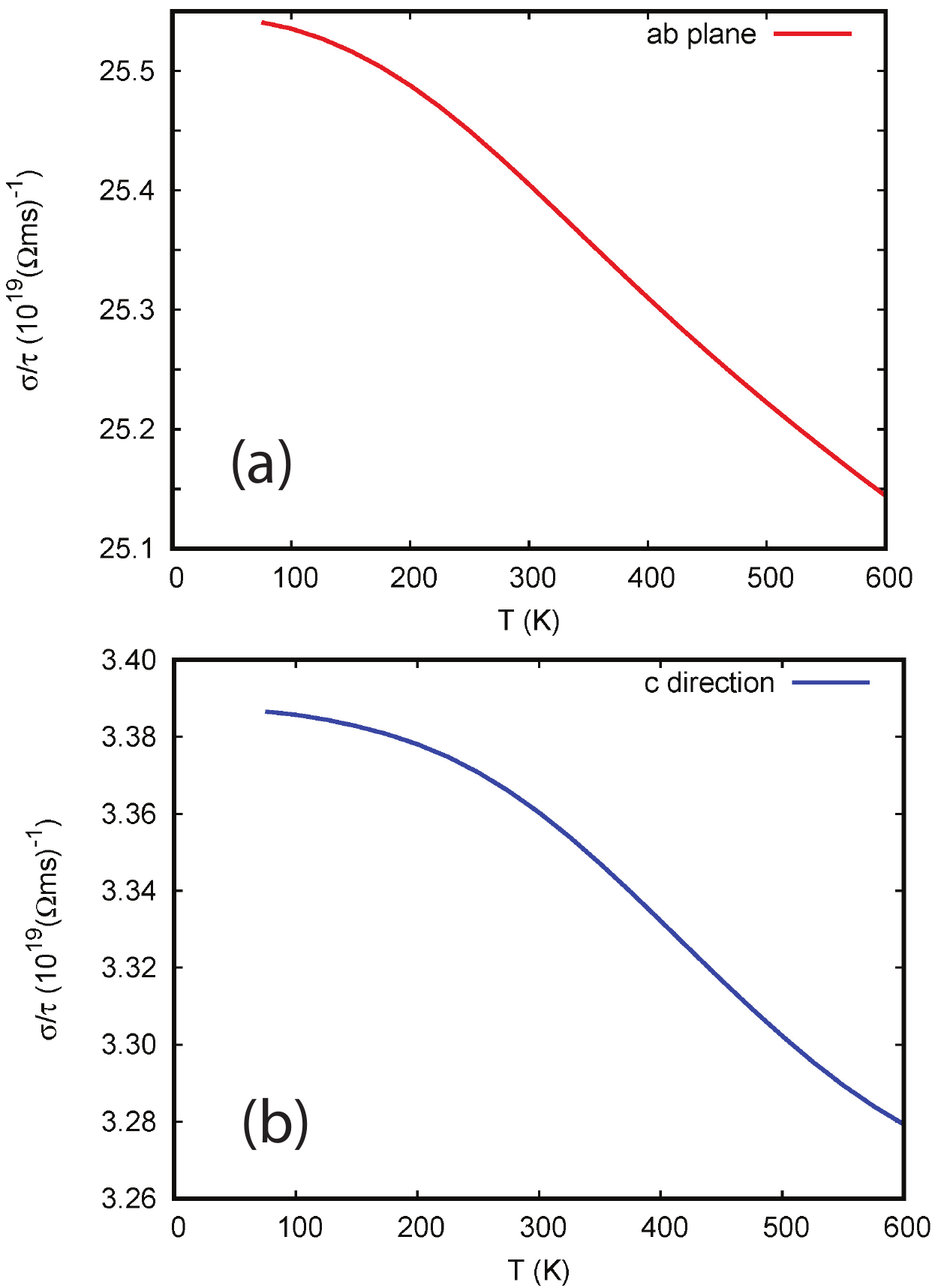}
\caption{Transport function of \kca\ versus temperature for electronic conduction (a)~in the $ab$~plane and (b)~along the $c$~axis.}
\label{Fig:xport_fcn}
\end{figure}

\begin{figure}
\includegraphics[width=3.3in]{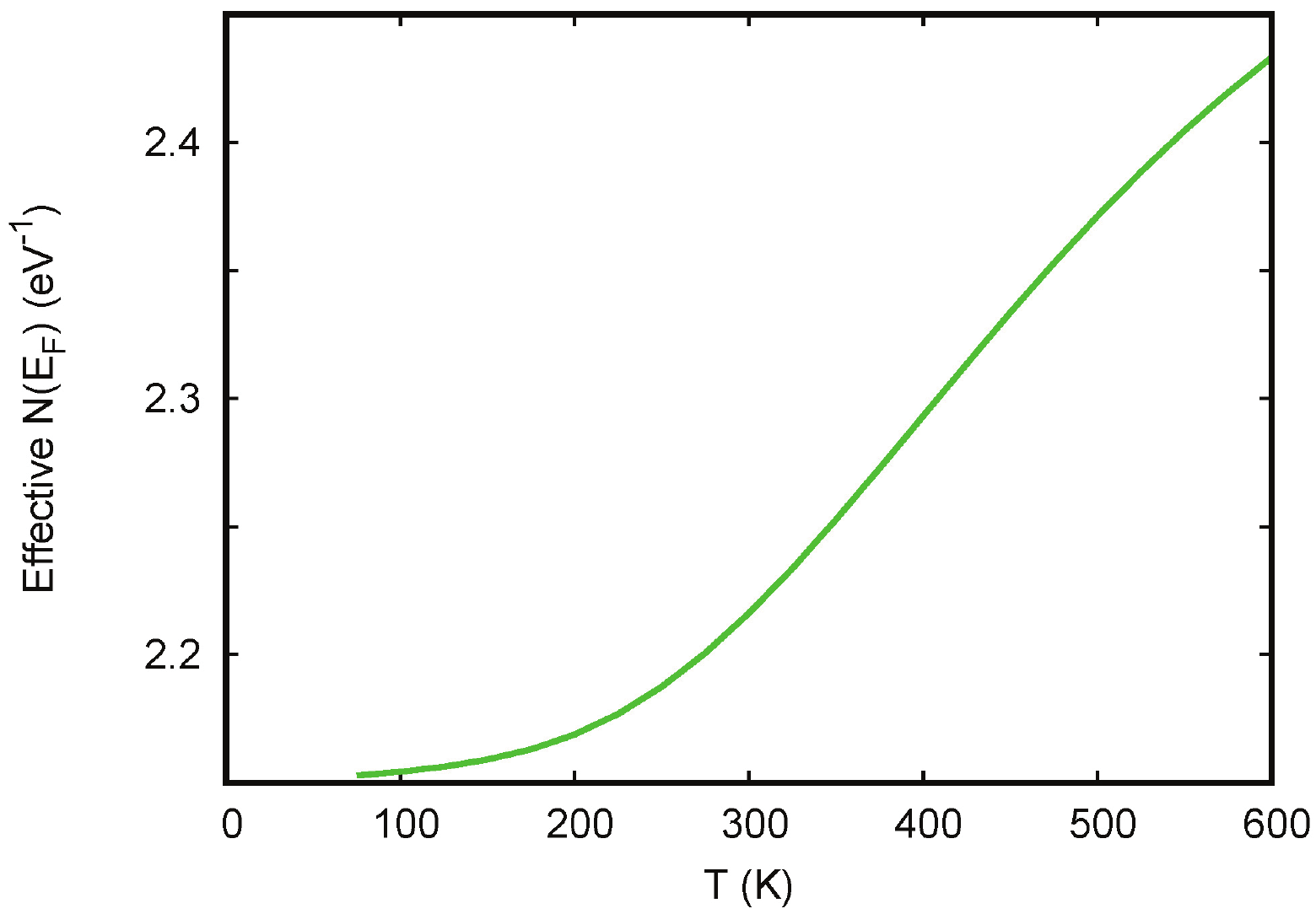}
\caption{Temperature dependence of the density of states at the Fermi level of \kca\ in units of states/eV\,f.u.\ for both spin directions.}
\label{Fig:Fermi_Level_vs_T}
\end{figure}

The transport calculations were carried out using BoltzTraP based on the first-principles electronic structure
\cite{boltztrap}. For this purpose we used eigenvalues on a dense 62$\times$62$\times$62 mesh in the Brillouin zone.
The electronic transport function is the quantity $\sigma/\tau$ where $\sigma$ is the electrical conductivity and $\tau$ is the mean electron-phonon scattering time.  The resistivity $\rho$ is then given by $\rho = [(\sigma/\tau)\tau]^{-1}$. Figures~\ref{Fig:xport_fcn}(a) and~\ref{Fig:xport_fcn}(b) show the transport function versus~$T$ for $ab$-plane and $c$-axis conduction, respectively. The calculated anisotropy is $\sigma_c/\sigma_{ab}=0.133$.  The behavior of the transport function with $T$ can contribute to the positive curvature in $\rho(T)$ observed in Fig.~\ref{Fig:RvsTfits}(a).
The corresponding low-$T$ plasma frequencies are $\hbar\omega_{\rm p}=3.53$~eV in the $ab$~plane and $\hbar\omega_{\rm p}=1.29$~eV along the $c$-axis direction.

The density of states near $E_{\rm F}$ in Fig.~\ref{Fig:DOS} strongly increases with energy.  Thus as $T$ increases and the Fermi function broadens, the Fermi level increases.  Figure~\ref{Fig:Fermi_Level_vs_T} shows the $T$ dependence of the Fermi level which can also contribute to the positive curvature of $\rho_{ab}(T)$. However, we find that the combined effects of the $T$ dependences of the transport function and the Fermi level are much too small to account for the magnitude of the positive curvature in $\rho_{ab}(T)$. However, the presence of high-frequency optic modes with substantial electron-phonon coupling may be responsible for the positive curvature~\cite{Takatsu2007}.

The electron-phonon contribution to the resistivity at high temperature can be written in terms of a coupling constant $\lambda_{\rm tr}$, which is expected to be similar to the value of $\lambda_\gamma$ for the specific heat \cite{allen}.
At high~$T$ the slope of the $T$-dependent electron-phonon-based resistivity can be written in terms of the transport function and $\lambda_{\rm tr}$ as~\cite{allen,dahal}
\begin{equation}
{{{\rm d}\rho}\over{{\rm d}T}} = (\sigma/\tau)^{-1}(2\pi\lambda_{\rm tr}k_{\rm B}/\hbar),
\end{equation}
where $k_{\rm B}$ is the Boltzmann constant and $\hbar$ is the reduced Planck constant. As mentioned, the resistivity is superlinear over the measured temperature range. However, if one uses the slope near 300~K, d$\rho$/d$T=0.23\,\mu\Omega$\,cm/K and the calculated in-plane transport function, one may estimate $\lambda_{\rm tr}=0.71$. This is significantly larger than the value $\lambda_{\gamma} = 0.36$ estimated from the specific-heat enhancement, indicating that there is additional temperature-dependent scattering. It may also be noted that there is a correction factor to the linear temperature dependence of the electron-phonon resistivity valid at high~$T$, given by
\begin{equation}
F_{\rm th}^{-1} = (1-\theta^2/12T^2)
\end{equation}
where $\theta^2=\hbar^2\langle\Omega^2\rangle/k_{\rm B}^2$ and $\langle\Omega^2\rangle$ is a weighted average-square phonon frequency. Thus, participation of high-frequency modes can lead to a superlinear dependence of the resistivity extending up to ambient temperature, but due to the functional form of $F_{\rm th}$ this reverts to linear as the temperature increases.

\begin{figure}
\includegraphics[width=3.5in]{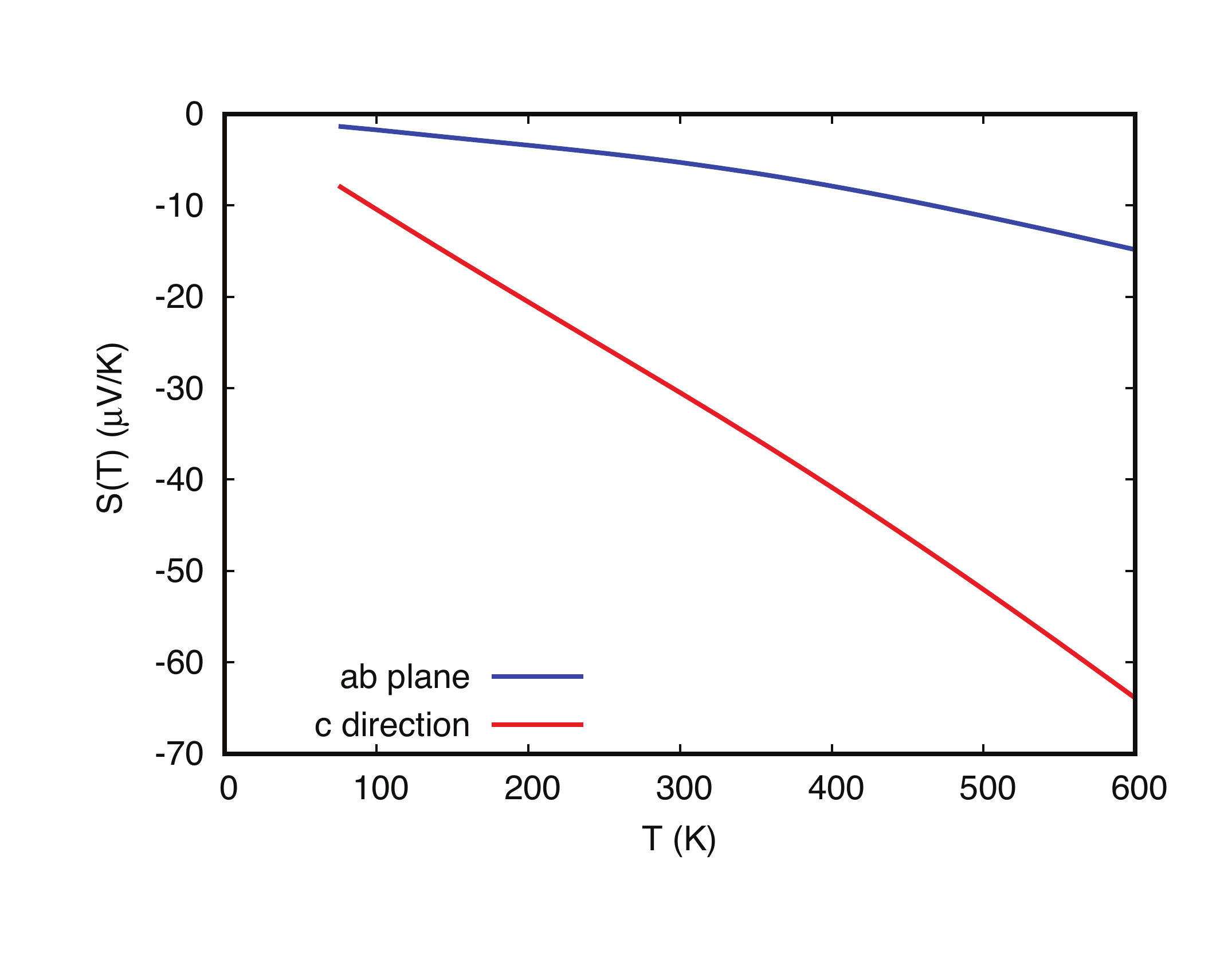}
\caption{Calculated temperature dependence of the anisotropic Seeback coefficient of \kca\ for heat trasport along the $ab$ plane and along the $c$~axis as indicated.}
\label{Fig:Thermopower}
\end{figure}

The calculated Seebeck coefficient shown in Fig.~\ref{Fig:Thermopower} is negative. This is consistent with the Fermi surfaces, which are electron-like. As mentioned, they consist of barrels at the zone corners, one from each of the two bands crossing the Fermi level, plus an electron cylinder at the zone center from the lower band and a smaller three-dimensional electron pocket on the $k_z$ face of the zone from the upper band. Interestingly, in contrast to most semiconductors and other materials with sizable Seebeck coefficients, there is a significant anisotropy as shown in Fig.~\ref{Fig:Thermopower}. Values at 300 K are $-5.3\,\mu$V/K in the $ab$~plane and $-30.5\,\mu$V/K along the $c$~axis. This is a consequence of the open Fermi surface sections. It would be interesting to measure this along with the $T$ dependence of the $c$-axis resistivity.

\begin{table*}
\caption{Comparison of the structure and property parameters of three previously reported Co-based 122-compounds $A$Co$_2$As$_2$ ($A =$ Ca, Sr, Ba) with those of KCo$_2$As$_2$. All these compounds crystallize in ThCr$_2$As$_2$-type tetragonal structure. Only CaCo$_{1.86}$As$_2$ exhibits long-range magnetic ordering a low temperatures. The listed structure parameters are $a$, $c$, and $z_{\rm As}$, and the distance between two nearest interlayer As atoms $d_{\rm As-As}$. The listed property parameters are the Sommerfeld coefficient $\gamma$, coefficient $\beta$ of the lattice heat capacity, the density of states (DOS) at the Fermi energy $E_{\rm F}$ derived from the $\gamma$ value ${\cal D}_\gamma(E_{\rm F})$, DOS at $E_{\rm F}$ obtained from band structure calculations ${\cal D}_{\rm band}(E_{\rm F})$, DOS at $E_{\rm F}$ estimated from magnetic susceptibility data ${\cal D}_\chi(E_{\rm F})$ and the Pauli formula, the Wilson ratio $R_{\rm W} = {\cal D}_\chi(E_{\rm F})/{\cal D}_\gamma(E_{\rm F})$, the Debye temperature $\Theta_{\rm D}$ estimated from heat capacity data, the residual resistivity ratio $\rho_{\rm 300\,K}/\rho_{\rm 2\,K}$, and the parameter $(1+\lambda_\gamma)m^\ast/m_{\rm band}$, where $\lambda_\gamma$ is a coupling constant and $m^\ast/m_{\rm band}$ is an electronic-mass enhancement. Magnetic ground states are abbreviated as: antiferromagnetic (AFM) or no long-range order (NLRO).}
\label{Table:Parameters}
 \begin{ruledtabular}
		\begin{tabular}{l c c c c}
		 Parameter & CaCo$_{1.86}$As$_2$ & SrCo$_2$As$_2$ & BaCo$_2$As$_2$  & KCo$_2$As$_2$ \\
		           & (Ref.~\cite{Anand-2014d}) & (Ref.~\cite{Pandey-2013b}) & (Ref.~\cite{Anand-2014a}) & (This work) \\ 
	 \hline
			\underline{Structure Parameters} &  &  &  & \\
			$a$(\AA) & 3.9837(2) & 3.9466(2) & 3.95743(2) & 3.813(1) \\
			$c$(\AA) & 10.2733(4) & 11.773(1) & 12.66956(9) & 13.58(1)\\
			$c/a$ & 2.5788(3) & 2.9831(4) & 3.20146(4) & 3.561(4)\\
			$z_{\rm As}$ & 0.3672(2) & 0.3587(3) & 0.35087(3)& 0.3491(2)\\
			$d_{\rm As-As}$(\AA) & 2.729(5) & 3.327(7) & 3.7788(8) & 4.098(9)\\
			\underline{Property Parameters} & & & & \\
			$\gamma$ (mJ/mol~K$^2$) & 27(1) & 37.8(1)  & 42(3) & 6.9(1)\\
			$\beta$ (mJ/mol~K$^4$)  & 1.00(8) & 0.610(7) & 0.36(4) & 0.514(9)\\
			${\cal D}_\gamma(E_{\rm F})$ & 11.4(5) & 16.0(3) & 18.82(2) & 2.92(5)\\
			${\cal D}_{\rm band}(E_{\rm F})$ [${\rm states/(eV\,f.u.)}$] & 2.95	& 11.04  & 8.23 & 2.15  \\
			${\cal D}_\chi(E_{\rm F})$ [${\rm states/(eV\,f.u.)}$] & --- & 54  & 90 & 5.54\footnotemark[2] \\
			$R_{\rm W}$ & --- & 3.4 & 5.2 & 1.88 \\
			$\Theta_{\rm D}$ (K) (Low $T$) & 212(1) & 251(1) & 302(12) & 266(2) \\
			$\Theta_{\rm D}$ (K) (All $T$) & 357(4) & 304(3)  & 306(9)  & 316(2) \\
			$\rho_{\rm 300\,K}/\rho_{\rm 2\,K}$ & 2.2\footnotemark[1] & 15.3 & 21.8 & 150 \\
			$(1+\lambda_\gamma)m^\ast/m_{\rm band}$ & 3.86 & 1.45  & 2.29 & 1.36(3)\\	
			Magnetic ground state & A-type AFM & NLRO & NLRO  & NLRO \\
		\end{tabular}
\footnotetext[1]{The ratio was taken between the $\rho$ values at $T = 300$ and 4~K\@.}
\footnotetext[2]{The value at 200~K from Table~\ref{Table:M(H)}.}
\end{ruledtabular}
\end{table*}

\subsection{\label{ARPES} Angle-resolved photoemission spectroscopy}

\begin{figure}
\includegraphics[width=3.3in]{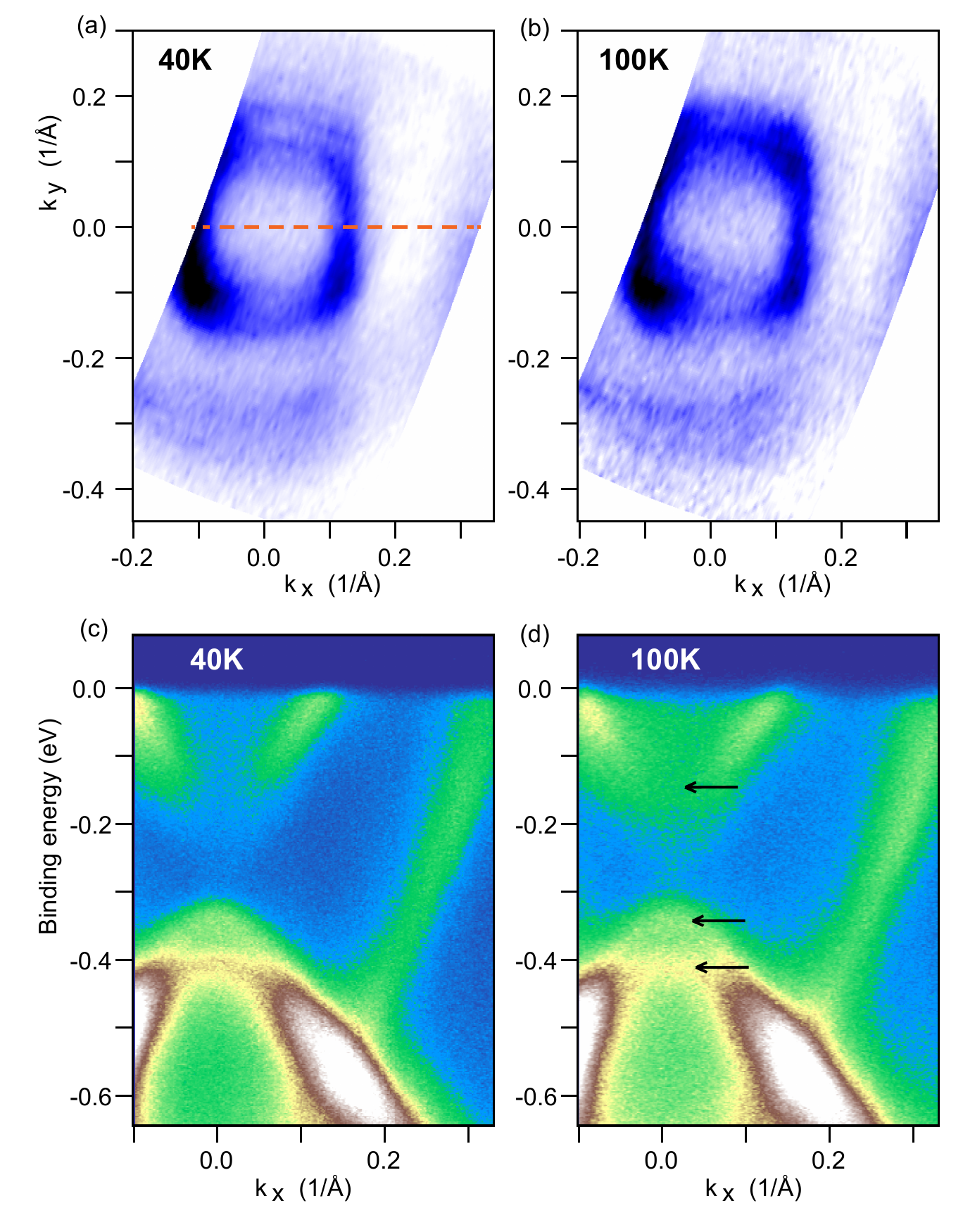}
\caption{Angle-resolved photoemission spectroscopy (ARPES) data on a KCo$_2$As$_2$ single crystal. (a) Intensity at the $E_{\rm F}$ integrated within 10~meV at $T = 40$~K. (b) Same as in the panel (a) but measured at $T = 100$~K. (c) Intensity along a momentum cut shown as red dashed line in the panel (a) measured at $T = 40$~K. (d) Same as in the panel (c), but measured at $T = 100$~K\@.  The arrows in the panel~(d) indicate energy locations of the band extrema of the three bands centered at~$\Gamma$.}
\label{fig:ARPES}
\end{figure}

The ARPES data revealed that the Fermi surfaces close to the $\Gamma$ point consist of round and concentric electron pockets [Figs.~\ref{fig:ARPES}(a) and (b)].  While the inner pocket is due to a band whose bottom is located at $-150$~meV with respect to the Fermi energy, the whole of the outer pocket arises from a band with the bottom located at about $-600$~meV [Figs.~\ref{fig:ARPES}(c) and (d)]. The inner band is separated from the top of the band below it by an energy gap of $\sim$200~meV\@. The band of the outer Fermi surface appears to cross this fully-occupied band without significant hybridization, indicating different orbital character. No significant change in the band structure or Fermi surface were observed versus temperature, apart from the expected thermal broadening of the bands and Fermi edge between 40 and 100~K\@. 

\section{\label{Discussion} Discussion}

The crystal growth of bct 122-type compound KCo$_2$As$_2$ using KAs self flux yielded high-quality crystals as evident by the observed low residual resistivity $\rho_{0}$ and the associated large residual resistance ratio with a value of 150.  These results are unprecedented for similar compounds with the ThCr$_2$Si$_2$ structure.  The $\rho_{ab}(T)$ data do not follow the Bloch-Gr\"uneisen model.  First, the $\rho_{ab}(T)$ data below  30~K follow a $T^4$ temperature dependence instead of the BG $T^5$ prediction. Second, the higher-$T$ data exhibit a very unusual positive curvature with a temperature exponent decreasing smoothly from 4 to 1.5 on heating from 30~K to 300~K instead of the linear dependence expected at high-$T$ from the BG model.  Our first-principles calculations indicate that the magnitude of the positive curvature in $\rho_{ab}(T)$ cannot be explained by the temperature dependences of the transport function and Fermi level.

Positive curvature in $\rho(T)$ over the high-$T$ range has also been consistently observed in the layered high-conductivity oxide delafossites such as PdCoO$_2$ and PtCoO$_2$ which also exhibit small values of the residual resistivity and large values of the RRR~\cite{Takatsu2007, Kushwaha2015, Nandi2018}. These compounds also show highly anisotropic Seebeck coefficients \cite{ong,yordanov}, as we also predict for KCo$_2$As$_2$. There, it appears that the only proposal to explain the above resistivity behavior was carrier scattering by optic phonons.  It is conceivable that this scattering mechanism can also explain the positive curvature in the high-$T$ resistivity of \kca.  First, our fit to the lattice heat capacity of \kca\ by the Debye theory for $C_{\rm p}(T)$ due to excitations of acoustic vibrations is poor.  Second, the $C_{\rm p}(T)$ is still increasing at 300~K, whereas $C_{\rm p}$ data for similar metallic isostructural compounds have usually leveled out by that temperature and the fits of $C_{\rm p}(T)$ by the Debye model plus a Sommerfeld electronic term below 300~K are typically quite good, thus suggesting the influence of optic vibrations on $C_{\rm p}(T)$ of \kca.

Another significant result obtained from the transport measurements on KCo$_2$As$_2$ is the observation of large magnetoresistance $(\Delta\rho/\rho)_{ab}$ as well as its linear variation with $H$, which is quite unusual for metallic systems, especially those not exhibiting long-range magnetic ordering \cite{Mizutani-2003}. The underlying mechanism of this observation invites further experimental and theoretical investigations.

The $\chi(T)$ as well as the $C_{\rm p}(T)$ and $\rho_{ab}(T)$ data do not show any evidence of long-range magnetic ordering in KCo$_2$As$_2$ down to 1.8~K\@. The Stoner enhancement discussed above is modest, consistent with the Wilson ratio from experiments. Moreover, the observed $\chi(T)$ of KCo$_2$As$_2$ is about one order of magnitude smaller than those of its Sr and Ba counterparts. This indicates that this material will likely not exhibit FM fluctuations similar to those reported in SrCo$_2$As$_2$~\cite{Li-2019}. High-pressure studies on this material would also be quite insightful as they would help to explore the effect of reduced CoAs interlayer distance on the magnetic, transport and electronic properties. The observations made on KCo$_2$As$_2$ suggest that while this material bears some similarity with the other $A$Co$_2$As$_2$ compounds, its properties are mostly quite different and should be further explored.

Similar to the isostructural compounds $A$Co$_2$As$_2$ ($A =$ Ca, Sr, Ba), the ARPES data of KCo$_2$As$_2$ show electron pockets at the center as well as at the corners of the Brillouin zone. However, in contrast to the non-circular rhombus-shaped pockets found in the former compounds, the pocket at the zone center is more circular in KCo$_2$As$_2$. The observed changes in the Fermi surface topology are likely a manifestation of the difference in electron counts between the former and the latter as well as the considerably-different interatomic distances within the $ab$-plane due to the significantly smaller $a$ lattice parameter of KCo$_2$As$_2$. These effects together are expected to result in substantially different bonding strengths as well as different hybridization states between the ions. 

Overall, the ARPES data shown in Fig.~\ref{fig:ARPES} agree reasonably well with the results of the band-structure calculations. There are three clearly-visible  bulk bands present in data in the proximity of the $\Gamma$ point. One of the bands crosses the Fermi energy giving rise to a circular electron pocket centered at $\Gamma$ and its bottom resides at $-150$\,meV\@. The tops of two fully-occupied bands are located at $\approx -350$~meV and $-400$~meV as indicated by arrows in panel (d) of Fig.~\ref{fig:ARPES}. The bottom of the electron pocket and the top of the fully-occupied band are separated by a gap of $\sim200$~meV\@. The calculation results shown in Fig.~\ref{Fig:BandStruct} predict a single electron band crossing $E_{\rm F}$ at $\Gamma$ with the bottom at $-280$~meV and the tops of the fully-occupied bands at $-380$~meV and $-460$~meV with a $\approx 100$~meV gap. The close agreement between the calculations and experimental data indicates that band renormalization effects are quite small in this material. The ARPES data show an additional larger electron pocket with a highly dispersive, very sharp band. Most likely this is a surface state that would not appear in the bulk band-structure calculation.

\section{\label{Conclusion} Concluding Remarks}

KAs self~flux was used for the growth of high quality single crystals of KCo$_2$As$_2$. The $ab$-plane electrical resistivity is unusual, exhibiting a $T^4$ behavior below 30~K and positive curvature up to 300~K\@. The magneto-transport data exhibit large magnetoresistance at low temperatures as well and show an unusual linear variation with applied magnetic field. No evidence of any phase transitions below 300~K is observed. The magnetic susceptibility values are about an order of magnitude smaller than for the related compounds SrCo$_2$As$_2$ and BaCo$_2$As$_2$. Our results reveal KCo$_2$As$_2$ to be an important material for studying the physics of high-purity layered metals. High-pressure studies of KCo$_2$As$_2$ will be important to investigate the effect of reduction of the interlayer distance between the adjacent CoAs layers on the magnetic and transport properties. Other studies likely to lead to interesting results include measurements of the $c$-axis resistivity and the anisotropy of the Seeback coefficient.

Note added: Recently, a preprint on the properties of \kca\ crystals appeared with some results similar to ours~\cite{Campbell-2020}.

\acknowledgments

The work at Ames Laboratory was supported by the U.S.~Department of Energy, Office of Basic Energy Sciences, Division of Materials Sciences and Engineering.  Ames Laboratory is operated for the U.S.\ Department of Energy by Iowa State University under Contract \mbox{No.\,DE-AC02-07CH11358}. Work at the University of Missouri was supported by the U.S.~Department of Energy, Office of Basic Energy Sciences, Division of Materials Sciences and Engineering, Award \mbox{No.\,DE-SC0019114}. AP acknowledges the support from Wits FRC, URC and National Research Foundation of South Africa. 

\appendix*
\section{\label{Cryst} Crystallography}

A table of crystallographic parameters for \kca\ is given below.
\begin{table}[h]
\caption{Structural information and refinement parameters for single-crystal KCo$_2$As$_2$, which crystallizes in the body-centered-tetragonal ThCr$_2$Si$_2$-type structure.}
\label{Table:Structure}
\begin{ruledtabular}
\begin{tabular}{lc} 
 
Parameter & Estimated value \\
 \hline
		
Formula weight (g)  &  306.8 \\

Space group & $I4/mmm$ \\

Formula units/unit cell ($Z$) & 2\\

Unit cell parameters (\AA) & $a = 3.813(1)$~\AA \\
					& $c = 13.58(1)$~\AA \\

Unit cell volume (\AA$^3$) & 197.3(1) \\

Density (g\,cm$^{-3}$) & 5.16 \\

Absorption coefficient (mm$^{-1}$) & 25.85 \\

Theta range (degree) & 3.0--29.138 \\

Index ranges 	& $-4 \leq h \leq 4$\\
			& $-5 \leq k \leq 5$\\
			& $-16 \leq l \leq 18$ \\

Reflections collected & 553 \\

Indep. reflections & 102 \\

Data/parameters & 102/8 \\

Goodness-of-fit on $F^2$ & 1.2 \\

Final R indices $[I > 2\sigma(I)]$ 	& $R_1 = 0.0574$\\
							& $wR_2 = 0.1220$ \\

R indices (all data) 	& $R_1 = 0.0898$\\
				& $wR_2 = 0.1373$ \\

Largest diff. peak \& hole (e \AA$^{-3}$) & 2.08 [1.3~\AA\ from Co]\\
								& $-1.97$ [0.11~\AA\ from As]\\

\end{tabular}
\end{ruledtabular}

\end{table}


\begin{thebibliography}{99}

\bibitem{Kamihara-2008} Y. Kamihara, T. Watanabe, M. Hirano, and H. Hosono, Iron-Based layered superconductor La[O$_{1-x}$F$_x$]FeAs ($x = 0.05$--0.12) with $T_{\rm c} = 26$~K, J. Am. Chem. Soc. {\bf 130}, 3296 (2008).

\bibitem{Wang-2008} C. Wang, L. Li, S. Chi, Z. Zhu, Z. Ren, Y. Li, Y. Wang, X. Lin, Y. Luo, S. Jiang, X. Xu, G. Cao, and Z. Xu, Thorium-doping-induced superconductivity up to 56~K in Gd$_{1-x}$Th$_x$FeAsO, Europhys. Lett. {\bf 83}, 67006 (2008).

\bibitem{Paglione-2010} J. Paglione and R. L. Greene, High-temperature superconductivity in iron-based materials, Nature Phys. {\bf 6}, 645 (2010).

\bibitem{Johnston-2010} D. C. Johnston, The puzzle of high temperature superconductivity in layered iron pnictides and chalcogenides, Adv. Phys. {\bf 59}, 803 (2010).

\bibitem{Stewart-2011} G. R. Stewart, Superconductivity in iron compounds, Rev. Mod. Phys. {\bf 83}, 1589 (2011).

\bibitem{Fernandes2022} R. M. Fernandes, A. I. Coldea, H. Ding, I. R. Fisher, P.~J. Hirschfeld, and G. Kotliar, Iron pnictides and chalcogenides: a new paradigm for superconductivity, Nature {\bf 601}, 35 (2022).

\bibitem{Anand-2014d} V. K. Anand, R. S. Dhaka, Y. Lee, B. N. Harmon, A. Kaminski, and D. C. Johnston, Physical properties of metallic antiferromagnetic CaCo$_{1.86}$As$_2$ single crystals, Phys. Rev. B {\bf 89}, 214409 (2014).

\bibitem{Anand-2014a} V. K. Anand, D. G. Quirinale, Y. Lee, B. N. Harmon, Y. Furukawa, V. V. Ogloblichev, A. Huq, D. L. Abernathy, P. W. Stephens, R. J. McQueeney, A. Kreyssig, A. I. Goldman, and D. C. Johnston, Crystallography and physical properties of BaCo$_2$As$_2$, Ba$_{0.94}$K$_{0.06}$Co$_2$As$_2$, and Ba$_{0.78}$K$_{0.22}$Co$_2$As$_2$, Phys. Rev. B {\bf 90}, 064517 (2014).

\bibitem{Pandey-2013b} A. Pandey, D. G. Quirinale, W. Jayasekara, A. Sapkota, M. G. Kim, R. S. Dhaka, Y. Lee, T. W. Heitmann, P. W. Stephens, V. Ogloblichev, A. Kreyssig, R. J. McQueeney, A. I. Goldman, Adam Kaminski, B. N. Harmon, Y. Furukawa, and D. C. Johnston, Crystallographic, electronic, thermal, and magnetic properties of single-crystal SrCo$_2$As$_2$, Phys. Rev. B {\bf 88}, 014526 (2013).

\bibitem{Jayasekara-2013} W. Jayasekara, Y. Lee, A. Pandey, G. S. Tucker, A. Sapkota, J. Lamsal, S. Calder, D. L. Abernathy, J. L. Niedziela, B. N. Harmon, A. Kreyssig, D. Vaknin, D. C. Johnston, A. I. Goldman, and R. J. McQueeney,  Stripe Antiferromagnetic Spin Fluctuations in SrCo$_2$As$_2$, Phys. Rev. Lett. {\bf 111}, 157001 (2013).

\bibitem{Li-2019} Y. Li, Z. Yin, Z. Liu, W. Wang, Z. Xu, Y. Song, L. Tian, Y. Huang, D. Shen, D. W. Abernathy, J. L. Niedziela, R. A Ewings, T. G. Perring, T. M. Pajerowski, M. Matsuda, P. Bourges, E. Mechthild, Y. Su, and P. Dai, Coexistence of Ferromagnetic and Stripe Antiferromagnetic Spin Fluctuations in SrCo$_2$As$_2$, Phys. Rev. Lett. {\bf 122}, 117204 (2019).

\bibitem{SMART-1996} SMART; Bruker AXS, Inc.: Madison, USA, (1996).

\bibitem{Blessing-1995} R. H. Blessing, An empirical correction for absorption anisotropy, Acta Crystallogr. A, {\bf 51}, 33 (1995).

\bibitem{SHELXTL-2000} SHELXTL; Bruker AXS, Inc.: Madison, USA, (2000).

\bibitem{pbe} J. P. Perdew, K. Burke, and M. Ernzerhof, Generalized Gradient Approximation Made Simple, Phys. Rev. Lett. {\bf 77}, 3865 (1996).

\bibitem{lapw} D. J. Singh and L. Nordstrom, {\em Planewaves, Pseudopotentials and the LAPW Method}, 2nd ed. (Springer, Berlin, 2006).

\bibitem{wien2k} P. Blaha, K. Schwarz, F. Tran, R. Laskowski, G. K. H. Madsen, and L. D. Marks, WIEN2k: An APW+lo program for calculating the properties of solids, J. Chem. Phys. {\bf 152}, 074101 (2020).

\bibitem{Rozsa-1981} S. R\'ozsa and H.-U. Schuster, Zur struktur von KFe$_2$As$_2$, KCo$_2$As$_2$, KRh$_2$As$_2$ und KRh$_2$P$_2$, Z. Naturforsch. B {\bf 36}, 1668 (1981).

\bibitem{Goetsch2012} R. J. Goetsch, V. K. Anand, A. Pandey, and D. C. Johnston, Structural, thermal, magnetic, and electronic transport properties of the LaNi$_2$(Ge$_{1-x}$P$_x)_2$ system, Phys. Rev. B {\bf 85}, 054517 (2012).

\bibitem{Park2014} C.-H. Park, N. Bonini, T. Sohier, G. Samsonidze, B. Kozinsky, M. Calandra, F. Mauri, and N. Marzari, Electron-Phonon Interactions and the Intrinsic Electrical Resistivity of Graphene, Nano Lett. {\bf 14}, 1113 (2014).

\bibitem{Gopal-1966} E. S. R. Gopal, \emph{Specific Heats at Low Temperatures}, (Plenum, New York, 1966).

\bibitem{brinkman} W. F. Brinkman and S. Engelsberg, Spin Fluctuation Contributions to the Specific Heat, Phys. Rev. {\bf 169}, 417 (1968).

\bibitem{mazin} I. I. Mazin and D. J. Singh, Ferromagnetic Spin Fluctuation Induced Superconductivity in Sr$_2$RuO$_4$, Phys. Rev. Lett. {\bf 79}, 733 (1997).

\bibitem{Johnston2020} D. C. Johnston, Thermodynamics of the nonrelativistic free-electron Fermi gas in one, two, and three dimensions from the degenerate to the nondegenerate temperature regime, arXiv: 2012.07663v2 (2020).

\bibitem{janak} J. F. Janak, Uniform susceptibilities of metallic elements, Phys. Rev. B {\bf 16}, 255 (1977).

\bibitem{Singh-2009} D. J. Singh, Properties of KCo$_2$As$_2$ and alloys with Fe and Ru: Density functional calculations, Phys. Rev. B {\bf 79}, 174520 (2009).

\bibitem{boltztrap} G. K. H. Madsen, and D. J. Singh, BoltzTraP. A code for calculating band-structure dependent quantities, Comput. Phys. Commun. {\bf 175}, 67 (2006).

\bibitem{Takatsu2007} H. Takatsu, S. Yonezawa, S. Mouri, S. Nakatsuji, K. Tanaka, and Y. Maeno, Roles of High-Frequency Optical Phonons in the Physical Properties of the Conductive Delafossite PdCoO$_2$, J. Phys. Soc. Jpn. {\bf 76}, 104701 (2007).

\bibitem{allen}  P. B. Allen, Empirical electron-phonon $\lambda$ values from resistivity of cubic metallic elements, Phys. Rev. B {\bf 36}, 2920 (1987).

\bibitem{dahal} A. Dahal, J. Gunasekera, L. Harringer, D. K. Singh, and D.J. Singh, Metallic nickel silicides: Experiments and theory for NiSi and first principles calculations for other phases, J. Alloys Compds. {\bf 672}, 110 (2016).

\bibitem{Kushwaha2015} P. Kushwaha, V. Sunko, P. J. W. Moll, L. Bawden, J. M. Riley, N. Nandi, H. Rosner, M. P. Schmidt, F. Arnold, E. Hassinger, T. K. Kim, M. Hoesch, A. P. Mackenzie, and P. D. C. King, Nearly free electrons in a $5d$ delafossite oxide metal, Sci. Adv. {\bf 1}, e1500692 (2015).

\bibitem{Nandi2018} N. Nandi, T. Scaffidi, P. Kushwaha, S. Khim, M. E. Barber, V. Sunko, F. Mazzola, P. D. C. King,  H. Rosner, P. J. W. Moll, M. K\"onig, J. E. Moore, S. Hartnoll, and A. P. Mackenzie, Unconventional magneto-transport in ultrapure PdCoO$_2$ and PtCoO$_2$, npj Quantum Mater. {\bf 3}, 66 (2018).

\bibitem{ong} K. P. Ong, D. J. Singh, and P. Wu, Unusual Transport and Strongly Anisotropic Thermopower in PtCoO$_2$ and PdCoO$_2$, Phys. Rev. Lett. {\bf 104}, 176601 (2010).

\bibitem{yordanov} P. Yordanov, W. Sigle, P. Kaya, M. E. Gruner, P. Pentcheva, B. Keimer, and H. U. Habermeier, Large thermopower anisotropy in PdCoO$_2$ films, Phys. Rev. Mater. {\bf 3}, 085403 (2019).

\bibitem{Mizutani-2003} U. Mizutani, \emph{Introduction to the electron theory of metals}, (Cambridge University Press, Cambridge, 2003).

\bibitem{Campbell-2020} D. J. Campbell, B. Wilfong, M. P. Zic, G. Levy, M. X. Na, T. M. Pedersen, S. Gorovikov, P. Y. Zavalij, S. Zhdanovich, A. Damascelli, E. E. Rodriguez, and J. Paglione, Physical properties and electronic structure of single-crystal KCo$_2$As$_2$, arXiv:2010.03447 (2020).

\end{thebibliography}
\end{document}